\documentstyle[referee,psfig]{jfm}
 
\let\realverbatim=\verbatim
\let\realendverbatim=\endverbatim
\renewcommand\verbatim{\par\addvspace{6pt plus 2pt minus 1pt}\realverbatim}
\renewcommand\endverbatim{\realendverbatim\addvspace{6pt plus 2pt minus 1pt}}

\ifCUPmtlplainloaded
\else
\fi
\ifCUPmtlplainloaded

\fi
\ifCUPmtlplainloaded
  \font\bit = mtmib10 at 10.5pt \skewchar\bit ='177  
\else
  \font\bit = cmmib10 \skewchar\bit ='177  
\fi
\ifCUPmtlplainloaded
\else
  \font\tenbmi=cmmib10 at 10pt  \skewchar\tenbmi ='177
  \font\sevenbmi=cmmib10 at 7pt \skewchar\sevenbmi ='177
  \font\fivebmi=cmmib10 at 5pt  \skewchar\fivebmi ='177

  \newfam\bmifam
  \textfont\bmifam=\tenbmi
  \scriptfont\bmifam=\sevenbmi
  \scriptscriptfont\bmifam=\fivebmi
  
\fi
\newsavebox{\thalfbox}
\sbox{\thalfbox}{$\textstyle\frac{1}{2}$}

\newsavebox{\shalfbox}
\sbox{\shalfbox}{$\scriptstyle\frac{1}{2}$}

\newsavebox{\squartbox}
\sbox{\squartbox}{$\frac{1}{4}$} 

\newsavebox{\etbox}
\sbox{\etbox}{\boldmath$\eta$}

\newsavebox{\astrutbox}
\sbox{\astrutbox}{\rule[-5pt]{0pt}{20pt}}

\ifnfsstwo

\fi
\ifnfssone
  \newmathalphabet{\mathit}
    \addtoversion{normal}{\mathit}{cmr}{m}{it}
    \addtoversion{bold}{\mathit}{cmr}{bx}{it}

\fi
\ifoldfss

\fi

\mathchardef\varLambda="0103

\ifCUPmtlplainloaded
\else
\fi
\ifCUPmtlplainloaded
  \let\bcdot=\undefined
  \NewSymbolFont{bldsym}{mtbsy10}{'60}
  \NewMathSymbol{\bcdot}{2}{bldsym}{01}
\else
  \font\tenbms=cmbsy10          \skewchar\tenbms ='60
  \font\sevenbms=cmbsy10 at 7pt \skewchar\sevenbms ='60
  \font\fivebms=cmbsy10 at 5pt  \skewchar\fivebms ='60

  \newfam\bmsfam
  \textfont\bmsfam=\tenbms
  \scriptfont\bmsfam=\sevenbms
  \scriptscriptfont\bmsfam=\fivebms

  \edef\bsy{\hexnumber\bmsfam}
  \mathchardef\bnabla="0\bsy72
  \mathchardef\bcdotsymbol="0\bsy01
  \def\bcdot{\,\bcdotsymbol\,}
\fi
\def\etal{\mbox{\it et al.\ }}

\title[The moving contact line hydrodynamics]
{A variational approach to the moving contact line hydrodynamics}

\author[T. Qian, X.-P. Wang and P. Sheng]%
{T\ls I\ls E\ls Z\ls H\ls E\ls N\ls G\ns Q\ls I\ls A\ls N$^1$,\ns
X\ls I\ls A\ls O\ls -\ls P\ls I\ls N\ls G\ns W\ls A\ls N\ls G$^1$
\break
\and P\ls I\ls N\ls G\ns S\ls H\ls E\ls N\ls G$^2$}

\affiliation{$^1$Department of Mathematics,
The Hong Kong University of Science and Technology,
Clear Water Bay, Kowloon, Hong Kong, China\\[\affilskip]
$^2$Department of Physics and Institute of Nano Science and Technology,
The Hong Kong University of Science and Technology,
Clear Water Bay, Kowloon, Hong Kong, China}

\pubyear{0000}
\volume{000}
\pagerange{000-000}
\date{?? and in revised form ??}
\begin{document}

\maketitle

\begin{abstract}
In immiscible two-phase flows, contact line denotes the intersection of 
the fluid-fluid interface with the solid wall.  When one fluid displaces 
the other, the contact line moves along the wall.  A classical problem 
in continuum hydrodynamics is the incompatibility between 
the moving contact line and the no-slip boundary condition, as the latter 
leads to a non-integrable singularity.  The recently discovered 
generalized Navier boundary condition (GNBC) offers an alternative to 
the no-slip boundary condition which can resolve the moving contact line 
conundrum. We present a variational derivation of the GNBC through 
the principle of minimum energy dissipation (entropy production), as
formulated by Onsager for small perturbations away from the equilibrium.  
Through numerical implementation of 
a continuum hydrodynamic model, it is demonstrated that the GNBC 
can quantitatively reproduce the moving contact line slip velocity profiles 
obtained from molecular dynamics simulations.  
In particular, the transition from complete slip 
at the moving contact line to near-zero slip far away 
is shown to be governed by a power-law partial slip regime, 
extending to mesoscopic length scales.  
The sharp (fluid-fluid) interface limit of the hydrodynamic model, 
together with some general implications of slip versus no-slip, are discussed.
\end{abstract}

\section{Introduction}\label{intro}

The no-slip boundary condition states that there can be no relative motion 
at the fluid-solid interface (Batchelor 1991). It is generally regarded as 
a cornerstone in continuum hydrodynamics, owing to its proven applicability 
in diverse fluid-flow problems. However, decades ago it was discovered 
that in immiscible two-phase flows, the moving contact line 
(MCL), defined as the intersection of the fluid-fluid interface with 
the solid wall, is incompatible with the no-slip boundary condition
(Moffatt 1964; Hua \& Scriven 1971; Dussan \& Davis 1974;
Dussan 1976; Dussan 1979; de Gennes 1985). 
As shown by \cite{rolling}, under usual hydrodynamic assumptions,
viz. incompressible Newtonian fluids, no-slip boundary condition, and
smooth, rigid, solid walls, there is a velocity discontinuity at the MCL, 
and the tangential force exerted by the fluids on the solid bounding 
surface in the vicinity of the MCL is infinite. This is the well-known 
contact-line singularity.
In the past two decades, it was shown through molecular dynamics (MD) 
simulations that near-complete slip indeed occurs at the MCL 
(Koplik, Banavar \& Willemsen 1988; Koplik, Banavar \& Willemsen 1989;
Thompson \& Robbins 1989; Thompson, Brinckerhoff \& Robbins 1993).  
This finding presented a conundrum for classical hydrodynamics, 
due to a lack of viable alternatives apart from ad hoc fixes.  
Furthermore, in the absence of a viable boundary condition which can 
reproduce the MD results, accurate continuum description of immiscible 
flows in the micro- or nanoscales remained an elusive goal.

Through analysis of extensive MD data, it was recently discovered that 
there is indeed a differential boundary condition, denoted the generalized 
Navier boundary condition (GNBC), which resolves the MCL conundrum 
(Qian, Wang \& Sheng 2003). 
Here we show that the GNBC can be derived variationally from the principle 
of minimum energy dissipation (Onsager 1931a and 1931b), 
and its implementation through the use of a Cahn-Hilliard (CH) free energy 
functional (Cahn \& Hilliard 1958) leads to quantitative predictions 
in excellent agreement with MD simulation results. In what follows, 
the MCL problem is briefly recapitulated in Sec. \ref{recapitulation}.  
The variational derivation of the GNBC in Sec. \ref{variation} is followed 
by a numerical demonstration of its consequences in Sec. \ref{comparison}. 
It is shown that the transition from near-complete slip at the MCL 
to near-zero slip (no-slip) far away from the MCL is not confined to 
a molecular-sized region around the MCL. 
Instead, the transition follows a power-law profile of partial slipping, 
extending to mesoscopic scales (Qian, Wang \& Sheng 2004).  
The sharp/diffuse (fluid-fluid) interface limits of our theory and  
their associated issues are described in Sec. \ref{scaling}.  
In Sec. \ref{remarks} we discuss some general implications of 
replacing the no-slip boundary condition, which can be regarded as 
an approximation to the GNBC in single-phase flow regions, 
by the more accurate GNBC.  In particular, it is argued that 
slip and partial slip boundary conditions offer the prospect of nanoscale 
interface engineering to ``tune'' the amount of slipping.  

\section{Recapitulation of the moving contact line problem}
\label{recapitulation}

\begin{figure}[ht]
\centerline{\psfig{figure=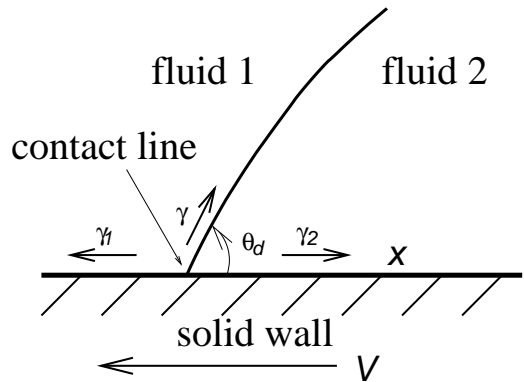,height=5.0cm}}
\caption{When one fluid displaces another immiscible fluid,
the contact line is moving relative to the solid wall. 
Due to the contact-line motion, the dynamic contact angle $\theta_d$
deviates from the static contact angle $\theta_s$ which is determined
by the Young equation $\gamma\cos\theta_s+\gamma_2=\gamma_1$.
}\label{CL}
\end{figure}

Consider an immiscible two-phase flow where one fluid displaces the other 
(see figure \ref{CL}).  If the no-slip boundary condition is applied along 
the wall, it can be shown that the tangential viscous stress varies as 
$\eta V/x$, where $\eta$ is the viscosity, $V$ is the wall speed in 
the reference frame where the fluid-fluid interface is stationary, and $x$ 
is the distance along the wall away from the MCL. This variation leads to 
diverging stress as $x\to 0$.  In particular, this stress divergence is 
non-integrable and implies infinite (viscous) dissipation 
(Dussan \& Davis 1974). 
Over the years there have been numerous models 
and proposals aiming to resolve this problem.  
For example, there have been the kinetic adsorption/desorption model 
by \cite{blake}, the slip models by \cite{hocking},
\cite{hua-mason}, and \cite{sheng}, 
and the diffuse-interface models by \cite{seppecher},
\cite{jacqmin}, \cite{vinal}, \cite{pismen-pomeau}, and \cite{yeomans}.

In another approach to the MCL problem, \cite{cox} carried out 
an asymptotic analysis and found, to the leading order in the capillary 
number $Ca=\eta V/\gamma$, a relation for the dependence of
the apparent contact angle (the angle of the fluid-fluid interface 
at some mesoscopic distance away from the fluid-solid interface) 
on the microscopic contact angle, the capillary number, and 
the distance from the contact line over which slip occurs. 
In this approach, the details of the slip flow in the inner region around 
the MCL are absorbed into a model-dependent constant in the asymptotic 
relation. In particular, it has been shown that the asymptotic behavior 
is independent of the microscopic boundary condition(s) (Dussan 1976).  
But this conclusion only deepens the mystery of what happens at
the contact line.

In the past two decades, MD simulations have shown that near-complete slip 
indeed occurs at the MCL 
(Koplik, Banavar \& Willemsen 1988; Koplik, Banavar \& Willemsen 1989;
Thompson \& Robbins 1989; Thompson, Brinckerhoff \& Robbins 1993).  
Small amount of partial slipping was also observed in single-phase flows
(Thompson \& Robbins 1990; Thompson \& Troian 1997;
Barrat \& Bocquet 1999a; Cieplak, Koplik \& Banavar 2001).  
Such partial slip can be accounted for 
by the Navier boundary condition (NBC), proposed nearly two centuries ago 
by \cite{navier1823}: $v^{slip}=l_s\dot{\gamma}$,
where $v^{slip}$ is the slip velocity at the surface, measured relative to 
the (moving) wall, $l_s$ is the slip length, and $\dot{\gamma}$ is 
the shear rate at the surface.
The small value of $l_s$ explains why the NBC is practically 
indistinguishable from the no-slip boundary condition in single-phase 
macroscopic flows.

While the NBC can account for the small amount of slip in high 
shear-rate single-phase flows, it fails to account, by an order of 
magnitude, for the near-complete slip at the MCL 
(Thompson \& Robbins 1989; Thompson, Brinckerhoff \& Robbins 1993).  
Recently, the MD-continuum hybrid simulation has been applied as 
a tool to investigate this problem by \cite{hadji}, and by \cite{weiqing}. 
But such approaches leave unresolved the problem of 
MCL boundary condition.  
Lack of a viable boundary condition implies that accurate MCL description 
(and hence immiscible two-phase flows) can only be attained through 
MD simulations, in systems far too small compared with most of  
the experimentally achievable samples.

The recent discovery of the GNBC (Qian, Wang \& Sheng 2003) 
resolves the MCL conundrum.  
The GNBC states that the slip velocity is proportional to 
the total tangential stress --- the sum of the viscous stress and 
the uncompensated Young stress; the latter arises from the deviation of 
the fluid-fluid interface from its static configuration.
Here we show the GNBC to be derivable from the principle of minimum energy 
dissipation. Its form is hence uniquely necessitated by the thermodynamics 
of immiscible two-phase flows.

\section{Variational derivation of the moving contact line hydrodynamics}
\label{variation}

There is a minimum dissipation theorem for incompressible single-phase 
flows. According to \cite{batchelor} (attributable to Helmholtz), 
``the rate of dissipation in the flow in a given region with negligible
inertia forces is less than that in any other solenoidal velocity
distribution (of zero divergence) in the same region with the same values 
of the velocity at all points of the boundary of the region.'' 
The variational principles involving dissipation have been further 
developed in the works of \cite{rayleigh}, \cite{onsager1}, 
\cite{onsager2}, \cite{edwards}, and \cite{doi}.
In mathematical terms, if we ignore inertia forces for the moment 
(this can be added in at the end, see equation (\ref{NSE})) and
let the variables 
$\alpha_1,\cdot\cdot\cdot,\alpha_n$ describe the displacement from 
thermodynamic equilibrium, 
$\dot{\alpha_1},\cdot\cdot\cdot,\dot{\alpha_n}$ being the corresponding 
rates and $F(\alpha_1,\cdot\cdot\cdot,\alpha_n)$ the free energy, then
for a set of simultaneous irreversible processes governed by 
\begin{equation}\label{onsager-1}
\sum_{j=1}^{n}\rho_{ij}\dot{\alpha}_j=-\displaystyle\frac{\partial F(\alpha_1,\cdot\cdot\cdot,\alpha_n)}{\partial \alpha_i},\;\;\;
(i=1,\cdot\cdot\cdot,n),
\end{equation}
where the coefficients $\rho_{ij}$ are introduced through 
the linear relations between the rates $\{\dot{\alpha_i}\}$ and 
the ``forces'' $\{-\partial F/\partial\alpha_i\}$,
a variational principle (Onsager 1931a and 1931b) can be formulated, namely
\begin{equation}\label{onsager-2}
\delta\left[\Phi(\dot{\alpha},\dot{\alpha})+
\dot{F}(\alpha,\dot{\alpha})\right]=\sum_{i=1}^{n}
\left(\displaystyle\frac{\partial\Phi}{\partial\dot{\alpha}_i}
+\displaystyle\frac{\partial F}{\partial \alpha_i}\right)
\delta\dot{\alpha}_i=0,
\end{equation}
where $\Phi(\dot{\alpha},\dot{\alpha})$ is the dissipation-function
defined by
\begin{equation}\label{onsager-3}
\Phi(\dot{\alpha},\dot{\alpha})\equiv \displaystyle\frac{1}{2}
\sum_{i,j}\rho_{ij}\dot{\alpha}_i\dot{\alpha}_j,
\end{equation}
and $\dot{F}(\alpha,\dot{\alpha})$ is the rate of change of 
the free energy:
\begin{equation}\label{onsager-4}
\dot{F}(\alpha,\dot{\alpha})\equiv \sum_{i=1}^{n}\displaystyle\frac
{\partial F(\alpha_1,\cdot\cdot\cdot,\alpha_n)}{\partial \alpha_i}
\dot{\alpha}_i.
\end{equation}
Microscopic reversibility requires the reciprocal relations 
$\rho_{ij}=\rho_{ji}$. We note that the variation 
in equation (\ref{onsager-2}) should be taken 
{\it with respect to the rates $\{\dot{\alpha_i}\}$ for prescribed 
$\{\alpha_i\}$}, and the extremum given by equation (\ref{onsager-2})
is always a minimum because $\Phi(\dot{\alpha},\dot{\alpha})$ is
quadratic in $\{\dot{\alpha_i}\}$ and positive definite (Onsager 1931b). 
The dissipation-function $\Phi$ as given by equation (\ref{onsager-3})
is noted to be half the rate of energy dissipation, 
owing to the required consistency between the ``force balance''
condition (\ref{onsager-1}), the minimum condition (\ref{onsager-2}), 
and the fact that the dissipative forces are linear in rates.
It can be shown that the principle of minimum energy dissipation
yields the most probable course of an irreversible process, provided
the displacements from thermodynamic equilibrium are small 
(Onsager 1931b, Onsager \& Machlup 1953).
We note that the variational principle presented here is a special form
of the principle of minimum energy dissipation formulated by Onsager.
First, the dissipation-function $\Phi$ defined here is different from
that by Onsager by a factor of $T$, the temperature which is assumed 
to be uniform in the fluids. Second, the rate of change of 
the free energy $\dot{F}$ equals $-T\dot{S}+\dot{W}$, where 
$S$ and $W$ denote the entropy and work, respectively (Doi 1983). 
This point will be elaborated after equation (\ref{rate-F-2}).
In Appendix \ref{appendix-a} we use a single-variable case to illustrate 
the underlying physics of the principle of minimum energy dissipation.

When applied to a single-phase flow confined by solid surfaces, 
the variational principle in equation (\ref{onsager-2}) becomes
\begin{equation}\label{onsager-helmholtz}
\delta\Phi(\dot{\alpha},\dot{\alpha})=\sum_{i=1}^{n}\displaystyle\frac
{\partial\Phi}{\partial\dot{\alpha}_i}\delta\dot{\alpha}_i=0.
\end{equation}
Here the rates $\{\dot{\alpha_i}\}$ correspond to the velocity field
${\bf v}({\bf r})$ and the $\dot{F}(\alpha,\dot{\alpha})$ term 
defined by equation (\ref{onsager-4}) drops out in the variation 
because there is no such free energy $F(\{\alpha_i\})$ 
in single-phase flows. As the dissipation-function 
$\Phi(\dot{\alpha},\dot{\alpha})$ equals half the rate of energy dissipation, equation (\ref{onsager-helmholtz}) leads directly to
the minimum dissipation theorem by Helmholtz (Batchelor 1991). That is, 
once the values of the velocity are prescribed at the solid surfaces, 
the rate of viscous dissipation is minimized by the solution of 
the Stokes equation.
Physically, when fluid slipping occurs at the solid surface, there is
dissipation at the fluid-solid interface as well. 
Below we show that to minimize the total rate of energy dissipation 
$R_1\left[{\bf v}\right]$ $(=2\Phi(\dot{\alpha},\dot{\alpha}))$, 
the incompressible flow has to satisfy the Stokes equation and the NBC
simultaneously. 
Mathematically, this means that the Stokes equation 
\begin{equation}\label{SE}
-\nabla p +\eta\nabla^2{\bf v}=0,
\end{equation}
and the NBC 
\begin{equation}\label{NBC}
\beta v_\tau^{slip}=-\sigma^{visc}_{n\tau}=
-\eta(\partial_n v_\tau+\partial_\tau v_n),
\end{equation}
can be derived by minimizing the functional 
\begin{equation}\label{single-functional}
R_1\left[{\bf v}\right]=\int d{\bf r}\left[\displaystyle\frac{\eta}{2}
(\partial_i v_j+\partial_j v_i)^2\right]+
\int dS\left[\beta\left(v_\tau^{slip}\right)^2\right],
\end{equation}
with respect to the velocity distribution.
Here $p$ is the pressure which plays the role of the Lagrange multiplier
for the incompressibility condition $\nabla\cdot{\bf v}=0$, $\eta$ is 
the shear viscosity, $\beta$ is the slip coefficient, subscript $n$ 
denotes the outward surface normal, subscript $\tau$ denotes 
the direction tangential to the surface, $v_\tau^{slip}$ is the slip 
velocity, defined as the tangential fluid velocity at the solid surface,
measured relative to the (moving) wall, $\sigma^{visc}_{n\tau}$ is
the $n\tau$ component of the Newtonian viscous stress tensor,
and $\int dS$ denotes the integration over the solid surface.
The functional $R_1[{\bf v}]$ is $2\Phi(\dot{\alpha},\dot{\alpha})$ 
for single-phase flows, measuring the total rate of dissipation 
due to viscosity in the bulk and slipping at the solid surface. 
We write the single-phase flow dissipation $R_1$ as the sum of $R_v$ 
due to viscosity and $R_s$ due to slipping: $R_1=R_{v}+R_{s}$, with 
\begin{equation}\label{RV}
R_{v}\left[{\bf v}\right]=
\int d{\bf r}\left[\displaystyle\frac{\eta}{2}
(\partial_i v_j+\partial_j v_i)^2\right]
\end{equation}
and 
\begin{equation}\label{RS}
R_s\left[{\bf v}\right]
=\int dS\left[\beta\left(v_\tau^{slip}\right)^2\right].
\end{equation}
While $R_s$ may appear to be very different from $R_v$ in form, 
yet in reality the two terms are very similar if it is realized that
$v_\tau^{slip}$ is just the tangential velocity difference/differential
at the fluid-solid interface. Hence $\beta$ is just the viscosity divided
by a length scale $l_s$, defined as the slip length.
Associated with the variation of the velocity field ${\bf v}({\bf r})\to
{\bf v}({\bf r})+\delta{\bf v}({\bf r})$, the change in $R_v$ is given by
\begin{equation}\label{var-Rv}
\delta R_v=-2\eta\int d{\bf r}\left[
\partial_j\left(\partial_j v_i+\partial_i v_j\right)\delta v_i\right]+
2\eta\int dS\left[\partial_n v_\tau\delta v_\tau+
\partial_\tau v_n\delta v_\tau\right],
\end{equation}
and that in $R_s$ given by
\begin{equation}\label{var-Rs}
\delta R_s=2\beta\int dS\left[v_\tau^{slip}\delta v_\tau\right].
\end{equation}
Imposing the incompressibility condition $\partial_i v_i=0$ by the use of 
a Lagrange multiplier $\alpha({\bf r})$ leads to one more term
$\int d{\bf r}\left[\alpha\partial_i v_i\right]$, whose variation is given by
\begin{equation}\label{var-constraint}
\delta \int d{\bf r}\left[\alpha\partial_i v_i\right]=
-\int d{\bf r}\left[\partial_i\alpha\delta v_i\right].
\end{equation}
Here the boundary condition $v_n=0$ has been used at the solid surface, 
where only $\delta v_\tau$ is allowed.
From equations (\ref{var-Rv}), (\ref{var-Rs}), and (\ref{var-constraint}),
we obtain the Euler-Lagrange equations
\begin{equation}\label{ELeqv}
-2\eta\partial_j\left(\partial_j v_i+\partial_i v_j\right)
-\partial_i\alpha=0
\end{equation}
in the bulk and 
\begin{equation}\label{ELeqs}
2\eta\left(\partial_n v_\tau+\partial_\tau v_n\right)+2\beta v_\tau^{slip}=0
\end{equation}
at the surface.
Note that equation (\ref{ELeqv}) is identical to the Stokes equation 
(\ref{SE}) with $\partial_j v_j=0$ and $\alpha=-2p$, and equation 
(\ref{ELeqs}) reduces to the NBC (\ref{NBC}). That equations 
(\ref{SE}) and (\ref{NBC}) can both be derived variationally from 
the minimization of $R_1$ is a generalization of the minimum dissipation 
theorem by taking into account fluid slipping at the solid surface.

It should be emphasized that while $R_s$ arises from the assumption of
fluid-solid interface slipping, there is no specification of how much 
slipping there should be. In other words, even an infinitesimal amount of 
interface slipping would lead to equations (\ref{SE}) and (\ref{NBC}).
In particular, the no-slip boundary condition is obtained in the limit of
$\beta\to\infty$, corresponding to a vanishing slip length $l_s=\eta/\beta$.
In the other limit, $\beta\to 0$, we would have only the first term 
on the right-hand side of equation (\ref{single-functional}), and 
$\sigma^{visc}_{n\tau}=0$ on the boundary. Thus the NBC interpolates 
between the zero tangential viscous stress limit and the no-slip limit.

To generalize the functional $\Phi(\dot{\alpha},\dot{\alpha})=
\frac{1}{2}R_1[{\bf v}]$ from single-phase to immiscible
two-phase flows, it is recognized that a free energy functional is 
required to stabilize the interface separating the two immiscible fluids. 
Hence the introduction of a Landau free energy functional 
${\cal F}[\phi({\bf r})]$ is a necessity, presumably of the form 
(Cahn \& Hilliard 1958; Bray 1994)
\begin{equation}\label{free-energy}
{\cal F}[\phi({\bf r})]=\int d{\bf r}
\left[\displaystyle\frac{K}{2}\left(\nabla\phi\right)^2+f(\phi)\right],
\end{equation}
where the potential $f(\phi)$ has a double-well structure.
Here the phase field $\phi({\bf r})$ measures the (conserved) composition 
locally defined by $\phi=(\rho_2-\rho_1)/(\rho_2+\rho_1)$, with 
$\rho_1$ and $\rho_2$ being the number densities of the two fluid species. 
We also introduce the interfacial free energy per unit area at 
the fluid-solid interface, $\gamma_{fs}(\phi)$, which is a function of 
the local composition. Two quantities $\mu$ and $L$ can be defined from 
the variation of the total free energy
$$F={\cal F}[\phi]+\int dS\left[\gamma_{fs}(\phi)\right],$$
that is,
\begin{equation}\label{free-energy-variation}
\delta\left\{{\cal F}[\phi]+\int dS\left[\gamma_{fs}(\phi)\right]\right\}
=\int d{\bf r}\left[\mu\delta\phi\right]+\int dS\left[L\delta\phi\right],
\end{equation}
in which $\mu=\delta {\cal F}/\delta \phi=-K\nabla^2\phi+
\partial f(\phi)/\partial\phi$ by definition is the chemical potential 
and $L$, given by $K\partial_n\phi+\partial\gamma_{fs}(\phi)/\partial\phi$,
is the corresponding quantity at the solid surface.
Minimizing the total free energy with respect to $\phi$ yields 
the equilibrium conditions $\mu=C$ in the bulk and $L=0$ at the surface, 
$C$ being a constant acting as the Lagrange multiplier for 
the conservation of $\phi$. It will be shown later that $L=0$ leads to 
the Young equation for the static contact angle.

From the equilibrium conditions derived above, we see that deviations
from the two-phase equilibrium may be measured by the ``forces'' 
$\nabla\mu$ in the bulk and $L$ at the fluid-solid interface. 
For small perturbations away from the equilibrium, 
the additional rate of dissipation $R_\phi$ arises from system responses 
that are linear in $\nabla\mu$ and $L$. 
Such responses are described by the diffusive current $\bf J$ 
in the bulk and the material time derivative of $\phi$ at the solid surface,
i.e., $\dot{\phi}=\partial\phi/\partial t+v_\tau\partial_\tau\phi$. 
The conservation of $\phi$ means that the diffusive current and 
the material time derivative of $\phi$ satisfy the continuity equation
\begin{equation}\label{continuity}
\dot{\phi}\equiv
\displaystyle\frac{\partial\phi}{\partial t}+{\bf v}\cdot\nabla\phi
=-\nabla\cdot{\bf J}.
\end{equation}
But at the fluid-solid interface, diffusive transport normal
to the interface is possible ($\partial_n J_n\ne 0$ in general), 
hence the interfacial $\phi$ is not conserved.
In anticipation of the relevant dynamics governing the conserved $\phi$
in the bulk and nonconserved interfacial $\phi$, 
$\bf J$ in the bulk and $\dot{\phi}$ at the solid surface are hence
the two additional rates associated with the coexistence of two phases.
The additional rate of dissipation $R_\phi$ due to the displacement 
from the two-phase equilibrium may be constructed as a functional
quadratic in the rates, $R_\phi=R_d+R_r$, where
\begin{equation}\label{RD0}
R_d=\int d{\bf r}\left[\displaystyle\frac{{\bf J}^2}{M}\right],
\end{equation}
and
\begin{equation}\label{RR0}
R_r=\int dS\left[\displaystyle\frac{\dot{\phi}^2}{\Gamma}\right],
\end{equation}
with $M$ and $\Gamma$ introduced as two phenomenological parameters.
Combining $R_\phi$ with $R_1$ in equation (\ref{single-functional}),
we obtain the dissipation-function $\Phi(\dot{\alpha},\dot{\alpha})$ 
for immiscible two-phase flows:
\begin{equation}\label{dissipation-function-1}
\Phi=\int d{\bf r}\left[\displaystyle\frac{\eta}{4}
(\partial_i v_j+\partial_j v_i)^2\right]+\int dS\left[
\displaystyle\frac{\beta}{2}\left(v_\tau^{slip}\right)^2\right]
+\int d{\bf r}\left[\displaystyle\frac{{\bf J}^2}{2M}\right]
+\int dS\left[\displaystyle\frac{\dot{\phi}^2}{2\Gamma}\right],
\end{equation}
which equals half the total rate of energy dissipation $R_2$ 
in two-phase flows, i.e.,
\begin{equation}\label{dissipation-function-2}
\Phi=\displaystyle\frac{1}{2}R_2=\displaystyle\frac{1}{2}(R_1+R_\phi)= 
\displaystyle\frac{1}{2}(R_v+R_s+R_d+R_r),
\end{equation} 
Now the viscosity $\eta$ and slip coefficient $\beta$ in equation
(\ref{dissipation-function-1}), respectively, are understood to take on
their respective values for the two immiscible fluids on two sides of 
the interface.
Note that the right-hand side of equation (\ref{dissipation-function-1}) 
consists of four terms, contributed by the four physically distinct 
sources of dissipation --- the shear viscosity in the bulk, 
the fluid slipping at the solid surface, the composition diffusion 
in the bulk, and the composition relaxation at the solid surface.
In addition, each term that contributes to $\Phi$ is positive definite 
and quadratic in a rate that arises from the displacement from 
the equilibrium and accounts for a particular source of dissipation.
This quadratic dependence follows the general rule governing entropy 
production in a thermodynamic process (Landau \& Lifshitz 1997), 
it directly arises from the linear response to a small perturbation 
away from the equilibrium.

The rate of change of the free energy, i.e., 
$\dot{F}(\alpha,\dot{\alpha})$ in equation (\ref{onsager-4}),
may be written as
\begin{equation}\label{rate-F-1}
\dot{F}=\int d{\bf r}\left[\mu\displaystyle\frac{\partial\phi}{\partial t}
\right]+\int dS\left[L\displaystyle\frac{\partial\phi}{\partial t}\right],
\end{equation}
in accordance with the variation of the total free energy in equation
(\ref{free-energy-variation}). Substituting 
${\partial\phi}/{\partial t}=\dot{\phi}-{\bf v}\cdot\nabla\phi$
into equation (\ref{rate-F-1}) and using
$\int d{\bf r}\left[\mu\dot{\phi}\right]=
\int d{\bf r}\left[-\mu\nabla\cdot{\bf J}\right]=
\int d{\bf r}\left[\nabla\mu\cdot{\bf J}\right]$
with $\int d{\bf r}\left[\nabla\cdot(\mu{\bf J})\right]
=\int dS\left[\mu J_n\right]=0$ because of the impermeability condition
$J_n=0$ at the solid surface, we obtain
\begin{equation}\label{rate-F-2}
\dot{F}=\int d{\bf r}\left[\nabla\mu\cdot{\bf J}-
\mu{\bf v}\cdot\nabla\phi\right]+
\int dS\left[L(\dot{\phi}-v_\tau\partial_\tau\phi)\right].
\end{equation}
Note that the laws of thermodynamics require $\dot{F}=-T\dot{S}+\dot{W}$,
where $S$ and $W$ denote the entropy and work, respectively. 
Here $F$ is the free energy associated with the composition field $\phi$, consequently the entropy part $-T\dot{S}$ must arise from 
the composition diffusion (in the bulk) and relaxation 
(at the fluid-solid interface) while the work rate $\dot{W}$ is due to 
the work done by the flow to the fluid-fluid interface. That is,
\begin{equation}\label{rate-F-2a}
-T\dot{S}=\int d{\bf r}\left[\nabla\mu\cdot{\bf J}\right]+
\int dS\left[L\dot{\phi}\right],
\end{equation}
and
\begin{equation}\label{rate-F-2b}
\dot{W}=\int d{\bf r}\left[-{\bf v}\cdot(\mu\nabla\phi)\right]+
\int dS\left[-v_\tau(L\partial_\tau\phi)\right].
\end{equation}
It will be seen that $\mu\nabla\phi$ and $L\partial_\tau\phi$ are
the ``elastic'' force/stress exerted by the interface to the flow.
It is clear that in the steady state $\dot{F}=0$ because the work is 
fully transformed into entropy, i.e., $\dot{W}=T\dot{S}$. 
This should be obvious since 
in the steady state the diffusive transport of the fluid-fluid interface 
is balanced by its kinematic transport by the flow, 
hence its free energy is invariant in the course of time.

Therefore, for immiscible two-phase flows the variational principle 
in equation (\ref{onsager-2}) may be expressed by using the functional
\begin{equation}\label{variation-function}
\begin{array}{ll}
\Phi+\dot{F}= & \int d{\bf r}\left[\displaystyle\frac{\eta}{4}
(\partial_i v_j+\partial_j v_i)^2\right]+\int dS\left[
\displaystyle\frac{\beta}{2}\left(v_\tau^{slip}\right)^2\right]
+\int d{\bf r}\left[\displaystyle\frac{{\bf J}^2}{2M}\right]
+\int dS\left[\displaystyle\frac{\dot{\phi}^2}{2\Gamma}\right]+\\
& \int d{\bf r}\left[\nabla\mu\cdot{\bf J}-
\mu{\bf v}\cdot\nabla\phi\right]+
\int dS\left[L(\dot{\phi}-v_\tau\partial_\tau\phi)\right].
\end{array}
\end{equation}
Based on equation (\ref{variation-function}), a hydrodynamic model
for the contact-line motion can be derived by minimizing 
$\Phi+\dot{F}$ with respect to the rates 
$\{{\bf v},\;{\bf J},\;\dot{\phi}\}$,
supplemented with the incompressibility condition $\nabla\cdot{\bf v}=0$.

As $\Phi$ is quadratic in $({\bf J},\dot{\phi})$ and $\dot{F}$
is linear in $({\bf J},\dot{\phi})$, the Euler-Lagrange equation 
with respect to $\bf J$ is given by
\begin{equation}\label{diffusive-current}
{\bf J}=-M\nabla\mu,
\end{equation}
where the parameter $M$ introduced in equation (\ref{RD0})
is seen to have the meaning of a mobility coefficient.
Substituting equation (\ref{diffusive-current}) into the continuity 
equation (\ref{continuity}) for $\phi$ 
yields the anticipated advection-diffusion equation
\begin{equation}\label{advection-diffusion}
\dot{\phi}=
\displaystyle\frac{\partial\phi}{\partial t}+{\bf v}\cdot\nabla\phi
=-\nabla\cdot{\bf J}=M\nabla^2\mu.
\end{equation}
Similarly, the corresponding Euler-Lagrange equation for minimizing 
$\Phi+\dot{F}$ with respect to $\dot{\phi}$ at the solid surface is
\begin{equation}\label{relaxation}
\dot{\phi}=
\displaystyle\frac{\partial\phi}{\partial t}+v_\tau\partial_\tau\phi
=-\Gamma  L(\phi).
\end{equation}
That is, at the fluid-solid interface, the relaxation dynamics of 
the interfacial $\phi$ is linear in $L(\phi)$, i.e., Allen-Cahn
dynamics for nonconserved quantities (Bray 1994).

Now we show that the Stokes equation with the capillary force,
\begin{equation}\label{stokes}
-\nabla p +\eta\nabla^2{\bf v}+\mu\nabla\phi=0,
\end{equation}
and the GNBC with the uncompensated Young stress,
\begin{equation}\label{gnbc}
\beta(\phi) v_\tau^{slip}=-\eta(\partial_n v_\tau+\partial_\tau v_n)
+L(\phi)\partial_\tau\phi,
\end{equation}
can be obtained by minimizing $\Phi+\dot{F}$
with respect to the fluid velocity. Here $\mu\nabla\phi$ is 
the capillary force density (Chella \& Vinals 1996;
Qian, Wang \& Sheng 2003),
$\beta(\phi)$ is the slip coefficient which may locally depend on 
the composition, and $L(\phi)\partial_\tau\phi$ is the uncompensated 
Young stress which vanishes in equilibrium (Qian, Wang \& Sheng 2003).
From equations (\ref{dissipation-function-1}), 
(\ref{dissipation-function-2}), (\ref{rate-F-2}) and 
(\ref{variation-function}), we see that the dependence of $\Phi+\dot{F}$ 
on the velocity comes from $\frac{1}{2}R_1=\frac{1}{2}(R_v+R_s)$ 
in $\Phi$ and $\int d{\bf r}\left[-\mu{\bf v}\cdot\nabla\phi\right]+
\int dS\left[-L v_\tau\partial_\tau\phi\right]$ in $\dot{F}$.
Consider a variation of the velocity field ${\bf v}({\bf r})\to 
{\bf v}({\bf r})+\delta{\bf v}({\bf r})$. The associated changes
in $R_v$ and $R_s$ are already given by equations (\ref{var-Rv}) and 
(\ref{var-Rs}), and those in $\dot{F}$ are given by
\begin{equation}\label{var-rate-F}
\delta \dot{F}=
\int d{\bf r}\left[-\mu\nabla\phi\cdot\delta{\bf v}\right]+
\int dS\left[-L\partial_\tau\phi\delta v_\tau\right]=
\int d{\bf r}\left[-\mu\partial_i\phi\delta v_i\right]+
\int dS\left[-L\partial_\tau\phi\delta v_\tau\right].
\end{equation}
Combining equations (\ref{var-Rv}), (\ref{var-Rs}), (\ref{var-constraint}),
and (\ref{var-rate-F}), we obtain the Euler-Lagrange equations
\begin{equation}\label{ELeqB}
-\eta\partial_j\left(\partial_j v_i+\partial_i v_j\right)
-\displaystyle\frac{1}{2}\partial_i\alpha-\mu\partial_i\phi=0
\end{equation}
in the bulk and 
\begin{equation}\label{ELeqS}
\eta\left(\partial_n v_\tau+\partial_\tau v_n\right)
+\beta v_\tau^{slip}-L(\phi)\partial_\tau\phi=0
\end{equation}
at the surface.
Note that equation (\ref{ELeqB}) is identical to 
the Stokes equation (\ref{stokes})
with $\partial_j v_j=0$ and $\alpha=-2p$, and equation (\ref{ELeqS}) 
reduces to the GNBC (\ref{gnbc}).
An important point of this derivation is that the uncompensated Young 
stress at the boundary (last term on the left side of equation 
(\ref{ELeqS})) must accompany the capillary force density in the bulk
(last term on the left side of equation (\ref{ELeqB})),
both being the ``elastic'' interfacial force.
Hence the uncompensated Young stress is simply the manifestation of 
the fluid-fluid interfacial tension at the solid boundary.

Once the free energies ${\cal F}[\phi]$ and $\gamma_{fs}(\phi)$ are fixed,
the contact-line motion (in the regime of small Reynolds number)
is fully determined by equations (\ref{advection-diffusion}), 
(\ref{relaxation}), (\ref{stokes}), and (\ref{gnbc}), 
supplemented by the incompressibility condition 
$\nabla\cdot{\bf v}=0$ and the impermeability conditions 
$v_n=0$ and $\partial_n\mu=0$ at the solid surface
(Qian, Wang \& Sheng 2003). The Stokes equation can be readily 
generalized to the Navier-Stokes equation
\begin{equation}\label{NSE}
\rho\left[{\partial{\bf v}\over
\partial t}+ \left({\bf v}\cdot\nabla\right){\bf v} \right]=
-\nabla p +\eta\nabla^2{\bf v}+\mu\nabla\phi,
\end{equation} 
by including the inertia forces, where $\rho$ is the mass density. Together, the Navier-Stokes equation (\ref{NSE}), the GNBC (\ref{gnbc}),
the advection-diffusion equation (\ref{advection-diffusion}), and
equation (\ref{relaxation}) for the relaxation of interfacial $\phi$, 
form a consistent hydrodynamic model for the contact-line motion in 
immiscible two-phase flows, first presented by \cite{qws}. 
It is now clear that our model is necessitated by more general considerations.

\section{Comparison between MD and continuum results}\label{comparison}

To demonstrate the physical validity of our model, numerical solutions have 
been obtained for direct comparison to the MD velocity and interfacial 
profiles. For this purpose, we make use of the CH free energy functional 
(Cahn \& Hilliard 1958)
\begin{equation}\label{CH-free-energy}
{\cal F}_{CH}[\phi({\bf r})]=\int d{\bf r}
\left[\displaystyle\frac{K}{2}\left(\nabla\phi\right)^2+\left(
-\displaystyle\frac{r}{2}\phi^2+\displaystyle\frac{u}{4}\phi^4\right)\right],
\end{equation}
to fix the form of $f(\phi)$ for ${\cal F}[\phi]$ 
in equation (\ref{free-energy}).
Here $K$, $r$, and $u$ are material parameters 
that can be determined from the interfacial thickness $\xi=\sqrt{K/r}$, 
the interfacial tension $\gamma=2\sqrt{2}r^2\xi/3u$, 
and the two homogeneous equilibrium phases $\phi_\pm=\pm\sqrt{r/u}=\pm 1$. 

The two coupled equations of motion are the advection-diffusion
equation for the phase field $\phi({\bf r})$ and
the Navier-Stokes equation in the presence of the capillary force
density:
\begin{equation}\label{he2}
\displaystyle\frac{\partial\phi}{\partial t}
+{\bf v}\cdot\nabla\phi=M\nabla^2\mu,
\end{equation}
\begin{equation}\label{he1}
\rho\left[{\partial{\bf v}\over
\partial t}+ \left({\bf v}\cdot\nabla\right){\bf v} \right]=
-\nabla p
+\nabla\cdot{\mbox{\boldmath$\sigma$}}^v+\mu\nabla\phi+{\bf f}_{e},
\end{equation}
together with the incompressibility condition $\nabla\cdot{\bf
v}=0$. Here $M$ is the mobility coefficient, 
$\mu=\delta {\cal F}_{CH}/\delta \phi$ 
is the chemical potential derived from the CH free energy functional 
${\cal F}_{CH}$, $\rho$ is the mass density of the fluid,
$p$ is the pressure, ${\mbox{\boldmath$\sigma$}}^v=
\eta\left[(\nabla {\bf v})+(\nabla {\bf v})^T\right]$ is the
Newtonian viscous stress tensor with $\eta$ being the viscosity,
$\mu\nabla\phi$ is the capillary force density, and ${\bf f}_{e}$ is
the external force. The boundary conditions at the solid surface are
the impermeability conditions $\partial_n \mu=0$, $v_n=0$, the
relaxational equation for surface $\phi$:
\begin{equation}\label{he4}
\displaystyle\frac{\partial\phi}{\partial t}+v_\tau\partial_\tau\phi=
-\Gamma L(\phi),
\end{equation}
and the GNBC in continuum differential form:
\begin{equation}\label{he3}
\beta(\phi) v_\tau^{slip}=-\eta(\partial_n v_\tau+\partial_\tau v_n)
+L(\phi)\partial_\tau\phi.
\end{equation}
Here $\tau$ denotes the direction tangent to the solid surface,
$n$ denotes the outward surface normal,
$\Gamma$ is a positive phenomenological parameter,
$L(\phi)=K\partial_n\phi+\partial\gamma_{fs}(\phi)/\partial\phi$
with $\gamma_{fs}(\phi)$ being the fluid-solid interfacial free energy
per unit area, $\beta(\phi)$ is the slip coefficient
which may locally depend on the local surface composition $\phi$,
and $L(\phi)\partial_\tau\phi$ is the uncompensated Young stress.
We use $\gamma_{fs}(\phi)=(\Delta\gamma_{fs}/2)\sin(\pi\phi/2)$
which is a smooth interpolation from
$\gamma_{fs}(\phi_-)=-\Delta\gamma_{fs}/2$
to $\gamma_{fs}(\phi_+)=\Delta\gamma_{fs}/2$. According to
the Young equation for the static contact angle $\theta_s$:
$\gamma_{fs}(\phi_+)+\gamma\cos\theta_s=\gamma_{fs}(\phi_-)$,
we have $\Delta\gamma_{fs}=-\gamma\cos\theta_s$.

To interpret physically the second term on the right-hand side of 
equation (\ref{he3}), let us consider a fluid-fluid interface that 
intersects the planar solid surface $z=0$ with a contact angle $\theta$ relative to the $x$ axis. For simplicity we assume the contact line 
to be straight and parallel to the $y$ axis, and
hence $\tau=x$. If the fluid-fluid interface is gently curved, then
$\int_{int}dx\left[(K\partial_n\phi)\partial_x\phi\right]=
\int_{int}d\phi(K\partial_m\phi)\cos\theta$,
where $\int_{int}dx$ denotes the integration across the fluid-fluid interface
along $x$ and $\partial_m$ means taking spatial derivative along 
the fluid-fluid interface normal $m$, with $\partial_n\phi\approx
\partial_m\phi\cos\theta$. As $\int_{int}d\phi(K\partial_m\phi)=\gamma$, 
we have $\int_{int}dx\left[(K\partial_n\phi)\partial_x\phi\right]
=\gamma\cos\theta$. In the equilibrium, $L(\phi)=0$, and hence 
$$\int_{int}dx\left[L(\phi)\partial_x\phi\right]=
\int_{int}dx\left[(K\partial_n\phi+\partial\gamma_{fs}/\partial\phi)
\partial_x\phi\right]$$ vanishes.
This leads directly to the Young equation 
\begin{equation}\label{YE}
\gamma\cos\theta_s+\gamma_{fs}(\phi_+)-\gamma_{fs}(\phi_-)=0,
\end{equation}
where $\theta_s$ is the static contact angle, $\gamma\cos\theta_s$ 
comes from $\int_{int}dx\left[(K\partial_n\phi)\partial_x\phi\right]$,
and $\gamma_{fs}(\phi_+)-\gamma_{fs}(\phi_-)$ from
$\int_{int}dx\left[(\partial\gamma_{fs}/\partial\phi)\partial_x\phi\right]$.
When the fluids are in motion, integrating the uncompensated Young stress
$L(\phi)\partial_x\phi$ across the fluid-fluid interface along $x$ yields
\begin{equation}\label{unYS}
\int_{int}dx\left[L(\phi)\partial_x\phi\right]
=\gamma\cos\theta_d+\gamma_{fs}(\phi_+)-\gamma_{fs}(\phi_-)=
\gamma(\cos\theta_d-\cos\theta_s),
\end{equation}
where $\theta_d$ is the dynamic contact angle. Equation (\ref{unYS}) 
implies that the uncompensated Young stress arises from the deviation 
of the fluid-fluid interface from its static configuration.
However, it must be pointed out that here the contact angle is 
the so-called ``microscopic contact angle.'' It differs from 
the ``apparent contact angle'' when the contact line is in motion.
In particular, the apparent contact angle can change rather sharply 
when the contact line begins to move. But $\theta_d$ can only vary linearly 
and relatively slowly with velocity.

From equation (\ref{unYS}) and the fact that for moderate flow rates 
the field $\phi$ for gently curved (fluid-fluid) interface relaxes to 
essentially the local stationary structure, we can write
\begin{equation}\label{unYS-local}
L(\phi)\partial_x\phi
=\gamma\cos\theta_d f(x)+\partial_x\gamma_{fs},
\end{equation}
where $f(x)$ is a function which peaks at $x_{CL}$, the contact line
center, and $\int dx f(x)=1$.
In particular, when the static contact angle is $90^\circ$ and 
$\gamma_{fs}$ is a constant, $f(x)$ can be well approximated by 
$(3/4\sqrt{2}\xi)\cosh^{-4}[(x-x_{CL})/\sqrt{2}\xi]$, derived from 
the stationary solution $\phi_e=\phi_+\tanh[(x-x_{CL})/\sqrt{2}\xi]$.
Thus the uncompensated Young stress is a peaked function centered at
the contact line 
(note that $\partial_x\gamma_{fs}$ is also peaked at the interfacial region). 
While the above expressions give a physical interpretation to 
the uncompensated Young stress, it is noted that the presence of 
$\cos\theta_d$ makes the expression awkward for the purpose of calculation
(as $\theta_d$ should be the outcome of the calculation). Thus in actual
computations the formulation in terms of $\phi$ is prefered.

We have carried out the MD-continuum comparison in such a way that 
virtually {\it no} adjustable parameter is involved in the model
calculation (Qian, Wang \& Sheng 2003). This is achieved as follows.
There are a total of nine material parameters in our model:
$\rho$, $\eta$, $\beta$, $\xi$, $\gamma$, $|\phi_\pm|$, $\theta_s$
$M$, and $\Gamma$. Among these parameters, the first seven are directly
measurable in MD simulations. As for $M$ and $\Gamma$, 
their values are fixed through an optimized MD-continuum 
comparison: one flow field from MD simulation is best fitted by that from
hydrodynamic model calculation with optimized $M$ and $\Gamma$ values,
although in our case the fitting is not very sensitive to the values of
$M$ and $\Gamma$. That is, fitting can be almost equally good 
if $M$ and $\Gamma$ deviate from the optimal values.
It will be shown in Sec. \ref{scaling} that such insensitivity is not
an accident, i.e., as long as the values of $M$ and $\Gamma$ are 
in the right range for the hydrodynamic model to reach 
the sharp interface limit, the continuum predictions should not be 
sensitive to these parameter values.

Once all the parameter values are determined, predictions from
our hydrodynamic model can be readily compared to the results from 
a series of MD simulations with different external conditions. 
The overall agreement is excellent, thus demonstrating the validity of 
the GNBC and the hydrodynamic model. We emphasize that the MD-continuum 
agreement has been achieved from the molecular-scale vicinity of
the contact line (Qian, Wang \& Sheng 2003) 
to far fields at the large scale (Qian, Wang \& Sheng 2004).

\begin{figure}[ht]
\centerline{\psfig{figure=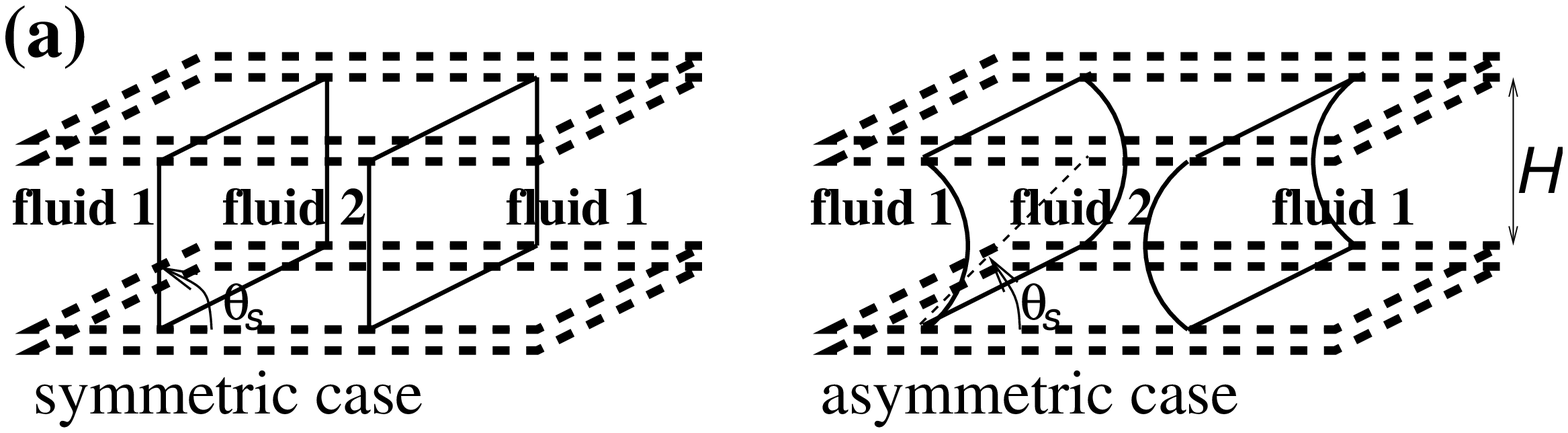,height=3.0cm}}
\bigskip
\centerline{\psfig{figure=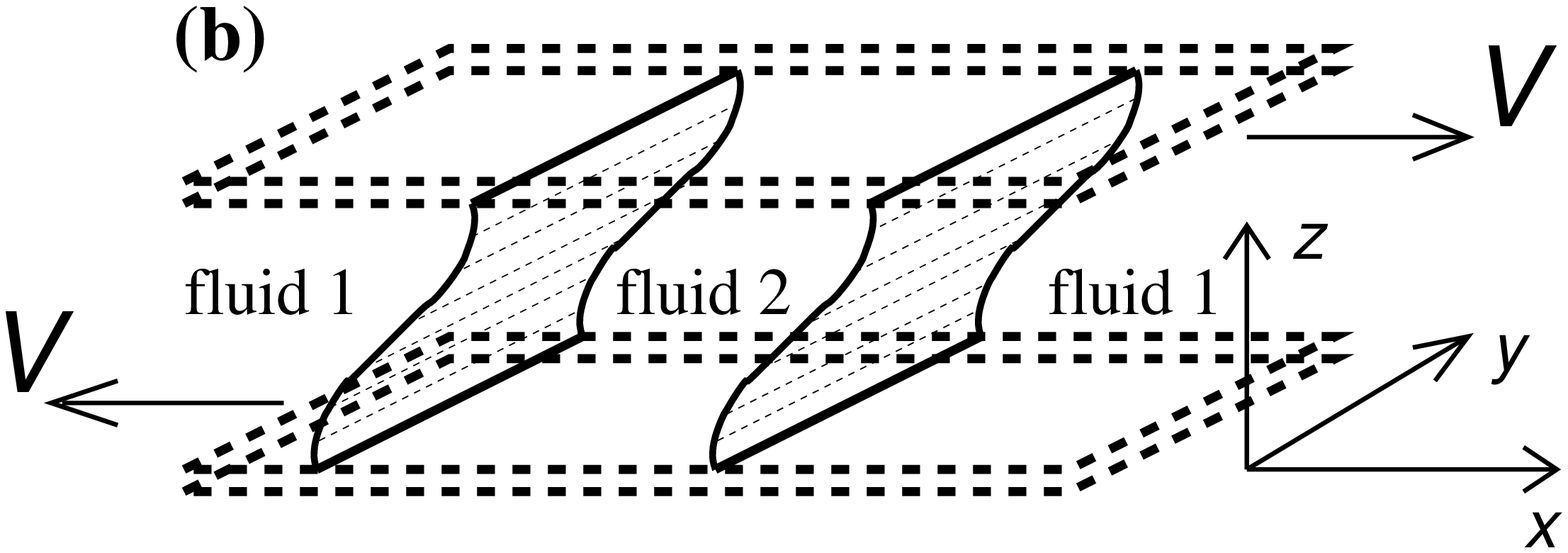,height=3.0cm}}
\caption{Schematic of simulation geometry.
(a) Static configurations in the symmetric and asymmetric cases. Fluid 2 
is sandwiched by fluid 1 due to the periodic boundary condition
along the $x$ direction. (b) Dynamic configuration in the symmetric case.
}\label{md-geometry}
\end{figure}

MD simulations have been carried out for immiscible two-phase flows in
Couette geometry (see figure \ref{md-geometry}) (Qian, Wang \& Sheng 2003). 
Two immiscible fluids were confined between two planar solid walls 
parallel to
the $xy$ plane, with the fluid-solid interfaces defined at $z=0$ and $H$. 
The Couette flow was generated by moving the top and bottom walls 
at a constant speed $V$ in the $\pm x$ directions, respectively. 
Periodic boundary conditions were imposed along the $x$ and $y$ directions.
Technical details of our MD simulations may be found from 
\cite{qws} and \cite{power-law}. 
Two cases were considered. In the symmetric case, 
the static contact angle $\theta_s$ is $90^\circ$ and 
the fluid-fluid interface 
is flat, parallel to the $yz$ plane. In the asymmetric case,
the static contact angle $\theta_s$ is $64^\circ$, 
and the fluid-fluid interface is curved in the $xz$ plane.
Steady-state velocity and interfacial profiles were obtained from 
time average over $10^5\tau$ or longer, where $\tau$ is the atomic time scale 
$\sqrt{m\sigma^2/\epsilon}$, with $\epsilon$ and $\sigma$ being the energy 
and length scales in the Lennard-Jones potential for fluid molecules,
and $m$ the fluid molecular mass. Throughout the remainder of this paper, 
all physical quantities are given in terms of the Lennard-Jones reduced units 
(defined in terms of $\epsilon$, $\sigma$, and $m$).

\begin{figure}
\centerline{\psfig{figure=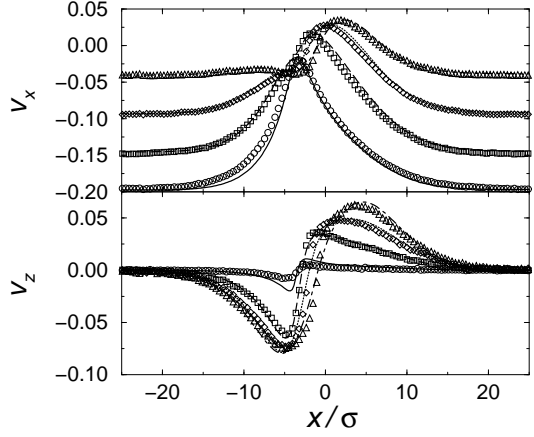,height=6.0cm}}
\caption{Comparison of the MD (symbols) and continuum (lines) 
velocity profiles ($v_x(x)$ and $v_z(x)$ at different $z$ levels)
for a symmetric case of immiscible Couette flow
($V=0.25(\epsilon/m)^{1/2}$ and $H=13.6\sigma$).
The profiles are symmetric about the center plane $z=H/2$,
hence only the lower half is shown at
$z=0.425\sigma$ (circles and solid lines), 
$2.125\sigma$ (squares and dashed lines),
$3.825\sigma$ (diamonds and dotted line), 
and $5.525\sigma$ (triangles and dot-dashed lines).
}\label{8symflow}
\end{figure}

\begin{figure}
\centerline{\psfig{figure=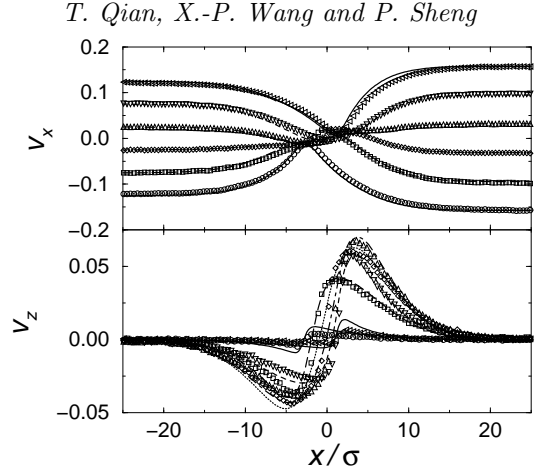,height=6.0cm}}
\caption{Comparison of the MD (symbols) and continuum (lines) 
velocity profiles ($v_x(x)$ and $v_z(x)$ at different $z$ levels)
for an asymmetric case of immiscible Couette flow
($V=0.2(\epsilon/m)^{1/2}$ and $H=13.6\sigma$), shown at
$z=0.425\sigma$ (circles and solid lines), 
$2.975\sigma$ (squares and long-dashed lines),
$5.525\sigma$ (diamonds and dotted line), 
$8.075\sigma$ (up-triangles and dot-dashed lines),
$10.625\sigma$ (down-triangles and dashed lines), 
$13.175\sigma$ (left-triangles and solid lines).
Although the solid lines are used to denote two different $z$ levels,
for each solid line, whether it should be compared to circles or
left-triangles is self-evident.
}\label{8asymflow}
\end{figure}

\begin{figure}
\centerline{\psfig{figure=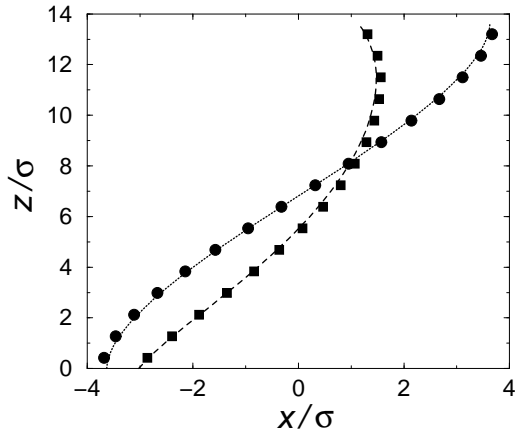,height=6.0cm}}
\caption{Comparison of the MD (symbols) and continuum (lines)
fluid-fluid interface profiles, defined by $\rho_1=\rho_2$ ($\phi=0$). 
The circles and dotted line denote the symmetric immiscible Couette flow 
with $V=0.25(\epsilon/m)^{1/2}$ and $H=13.6\sigma$; 
the squares and dashed line denote the asymmetric immiscible Couette flow 
with $V=0.2(\epsilon/m)^{1/2}$ and $H=13.6\sigma$.
}\label{interface8}
\end{figure}

Figure \ref{8symflow} shows the MD and continuum velocity fields for 
a symmetric case of Couette flow, figure \ref{8asymflow} shows those fields 
for an asymmetric case of Couette flow, and figure \ref{interface8} shows 
the MD and continuum fluid-fluid interface profiles for the above two cases,
from which the difference between the ``apparent'' advancing and receding
contact angles is clearly seen.

\begin{figure}
\centerline{\psfig{figure=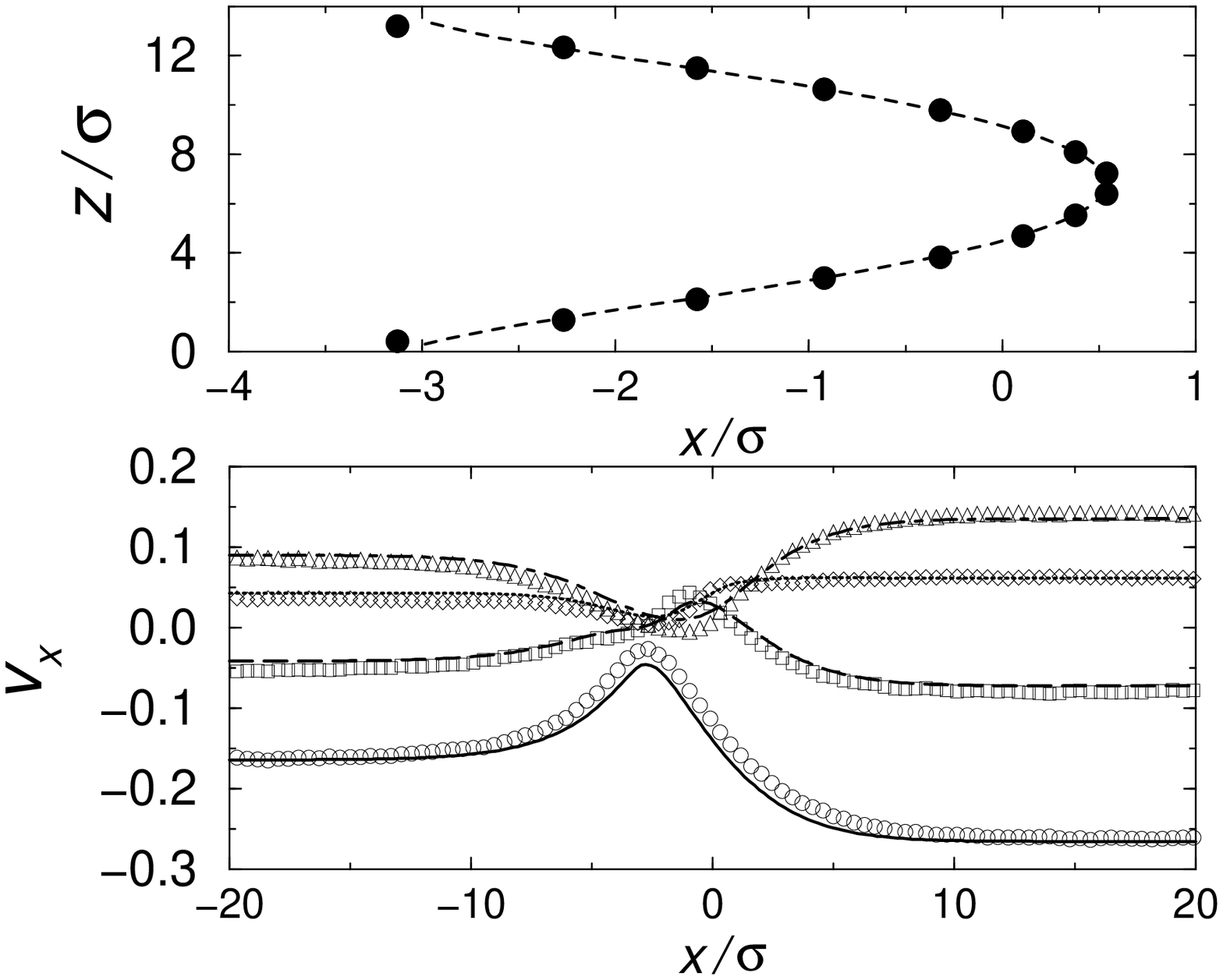,height=6.0cm}}
\caption{Comparison of the MD (symbols) and continuum (lines)
results for an asymmetric case of immiscible Poiseuille flow.
An external force $mg_{ext}=0.05\epsilon/\sigma$ is
applied on each fluid molecule in the $x$ direction,
and the two walls, separated by $H=13.6\sigma$, move at a constant speed 
$V=0.51(\epsilon/m)^{1/2}$ in the $-x$ direction to maintain 
a stationary steady-state interface. Upper panel: 
Fluid-fluid interface profiles, defined by $\rho_1=\rho_2$ ($\phi=0$). 
Lower panel: $v_x(x)$ at different $z$ levels.
The profiles are symmetric about the center plane $z=H/2$,
hence only the lower half is shown at
$z=0.425\sigma$ (circles and solid line), 
$2.125\sigma$ (squares and dashed line),
$3.825\sigma$ (diamonds and dotted line), 
and $5.525\sigma$ (triangles and dot-dashed line).
}\label{psf}
\end{figure}

In order to further verify that the hydrodynamic model is local and 
the parameter values are local properties, hence applicable to
different flow geometries, we have carried out MD and
continuum simulations for immiscible Poiseuille flows 
(Qian, Wang \& Sheng 2003). 
We find that the hydrodynamic model with the same set of parameters is 
capable of reproducing the MD velocity and interfacial profiles, 
shown in figure \ref{psf}. Similar to what's observed in Couette flows, 
here the slip is near-complete at the MCL, i.e., 
$v_x\approx 0$ and $|v_x^{slip}|\approx V$ (the wall speed),
while far away from the contact line, the flow field is not perturbed 
by the fluid-fluid interface and the single-fluid unidirectional 
Poiseuille flow is recovered.

The parameter values used for the MD-continuum comparison have been given
by \cite{qws}. We emphasize that the overall agreement is excellent 
in all cases, therefore the validity of the GNBC and the hydrodynamic model 
is well affirmed.

It should be mentioned that our MD results also show some fluid-fluid
interfacial structure which is outside the realm of our continuum model.
In particular, the MD data show a fairly significant density drop right
at the interface ($\phi=0$). Physically this is due to the repulsive 
molecular interaction between the two immiscible fluids. This density drop 
has implication on the slip coefficient $\beta$ in that for the interfacial 
region $\beta$ is no longer a simple composition of $\beta_1$ and $\beta_2$ 
on the two sides of the contact line.
At the same time, MD data also revealed and verified the form and nature 
of the uncompensated Young stress, just as predicted by equations 
(\ref{unYS}) and (\ref{unYS-local}). Details can be found in \cite{qws}.

Another important comparison between MD and continuum hydrodynamics is
the behavior that interpolates between the MCL, where there is near-complete
slip, and far away from the contact line, where there is at most 
a small amount of partial slip.

\begin{figure}
\centerline{\psfig{figure=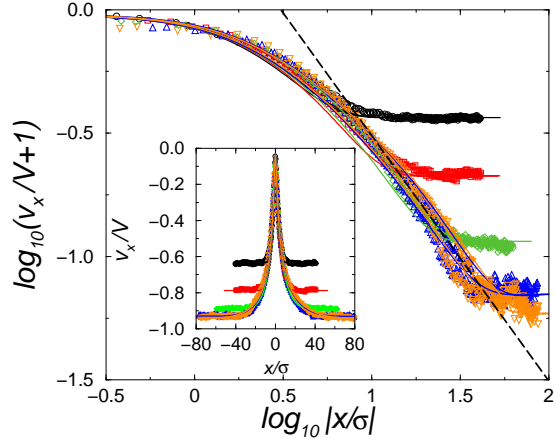,height=6.0cm}}
\caption{Log-log plot for the slip profiles. Here $v_x/V+1$ is 
the scaled slip velocity at the lower fluid-solid interface $z=0$, 
and $x/\sigma$ measures the distance from the MCL in units of $\sigma$. 
The wall is moving at $-V$, hence $v_x/V=0$ means complete slip and 
$v_x/V=-1$ means no slip. The $v_x$ profiles were obtained for 
five symmetric cases of Couette flow, which used different values for $H$ 
but the same value for $V$ ($=0.05\sqrt{\epsilon/m}$) and also 
the same parameters for densities and interactions.
The symbols represent the MD results and the solid lines represent 
the continuum results, obtained for 
$H=6.8\sigma$ (black circles and line), 
$H=13.6\sigma$ (red squares and line), 
$H=27.2\sigma$ (green diamonds and line), 
$H=54.4\sigma$ (blue up-triangles and line), 
$H=68\sigma$ (orange down-triangles and line).
There are two solid curves for each color, one for the slip profile 
left to the MCL and the other for that right to the MCL. 
The dashed line has the slope of $-1$, indicating that the $1/x$ behavior 
is approached for increasingly larger $H$. For $H=68\sigma$, 
the $1/x$ behavior extends from $|x|\approx 12\sigma\approx 6l_s$ to 
$50\sigma\approx 25l_s$, where $l_s$ was measured to be $2\sigma$. 
Inset: The scaled tangential velocity $v_x/V$ at $z=0$,
plotted as a function of $x/\sigma$.}\label{power-md}
\end{figure}

MD simulations have been carried out for immiscible Couette flows
in increasingly wider channels (Qian, Wang \& Sheng 2004). 
The inset to figure \ref{power-md} shows
the tangential velocity profiles at the wall. 
Immediately next to the MCL, there is a small core region,
where the slip profiles show a sharp decay within a few $l_s$ 
(the slip length in single-phase flows). As the channel width 
$H$ increases, a much more gentle variation of the slip profiles 
becomes apparent. In order to reveal the nature of this slow variation, 
we plot in figure \ref{power-md} the same data in the log-log scale. 
The dashed line has a slope $-1$, indicating the $1/x$ behavior of 
the slip profile, where $x$ is the distance from the MCL. 
Because of the finite $H$, there is always a plateau in each of 
the single-phase flow regions, where the constant small amount of slip 
is given by $v_0^{slip}=2Vl_s/(H+2l_s)$, which acts as an outer cutoff on 
the $1/x$ profile. This expression for $v_0^{slip}$ is simply derived from 
the NBC and the Navier-Stokes equation for uniform shear flow.
Our largest MD simulation shows that the $v^{slip}\propto 1/x$ behavior 
extends to $\sim 50\sigma$ (or $\sim 25l_s$). Obviously, 
as $H\rightarrow\infty$ and $v_0^{slip}$ approaches $0$ (no-slip), 
the power-law region can be very wide indeed. This large $1/x$ 
partial slip region indicates that the outer cutoff length scale 
(e.g., the system dimension) would determine the integrated effects, 
such as the total steady-state dissipation.  
Actually, in the past the similarity solutions of Stokes equation
have shown the $1/x$ stress variation away from the MCL 
(Moffatt 1964; Hua \& Scriven 1971). However, to our knowledge the fact 
that the partial slip also exhibits the same spatial dependence was 
first determined by \cite{power-law}, even though the validity of 
the NBC has been verified at high shear stress (Thompson \& Troian 1997;
Barrat \& Bocquet 1999a; Cieplak, Koplik \& Banavar 2001).

The continuum results shown in figure \ref{power-md} were obtained 
on a uniform mesh, using the same set of parameter values corresponding to 
the same local properties in all the five MD simulations.
We have extended the MCL simulations, through continuum hydrodynamics, 
to lower flow rates and much larger systems.  
For this purpose, we have employed the adaptive method based on iterative 
grid redistribution (Ren \& Wang 2000). The computational mesh is 
redistributed following the behavior of the continuum solution so that 
fine molecular resolution is achieved in the interfacial region, 
while elsewhere a much coarser mesh is used to save computational cost. 
A semi-implicit time stepping scheme is also used to speed up 
the approach to steady state.

\begin{figure}
\centerline{\psfig{figure=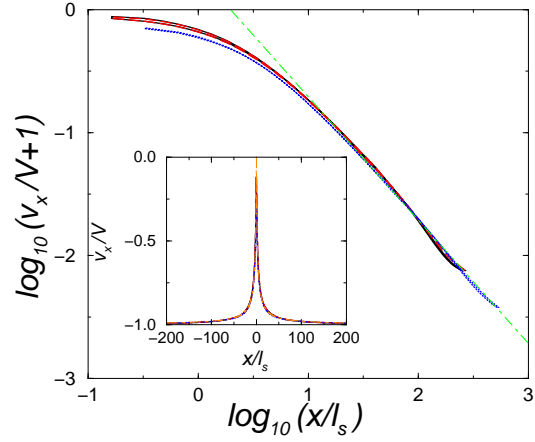,height=6.0cm}}
\caption{Log-log plot of the slip profiles. 
Here $v_x/V+1$ is the scaled slip velocity at the lower fluid-solid 
interface $z=0$, and $x/l_s$ measures the distance from the MCL 
in unit of $l_s$. The wall is moving at $-V$, hence $v_x/V=0$ means 
complete slip and $v_x/V=-1$ means no slip. 
There are three symmetric cases of Couette flow.
The black solid line denotes the case of 
$H=326\sigma$, $V=0.005\sqrt{\epsilon/m}$ and $l_s=1.24\sigma$, 
the red dashed line denotes the case of
$H=326\sigma$, $V=0.0025\sqrt{\epsilon/m}$ and $l_s=1.24\sigma$, 
and the blue dotted line denotes the case of 
$H=326\sigma$, $V=0.0025\sqrt{\epsilon/m}$ and $l_s=0.62\sigma$.
The green dot-dashed line has the slope of $-1$, indicating a $1/x$ region
much wider than that in figure \ref{power-md}. Inset: Universal slip profile. 
The scaled tangential velocity $v_x/V$ at $z=0$ is plotted as a function of 
the scaled coordinate $x/l_s$ for the three cases. 
The slip profiles exhibit a partial-slip region as large as hundreds of 
$l_s$. The relation $v_x/V=1/(1+{|x|}/{2.14l_s})-1$ 
is also plotted by the orange dot-dashed line, showing an extremely
good fit.}\label{power-hyd}
\end{figure}

The continuum results for three large systems of low flow rates 
are shown in figure \ref{power-hyd}. The capillary force is verified to be
important in the interfacial region only, whereas the pressure gradients 
and viscous forces show a much slower variation. They are balanced outside 
the interfacial region, and hence the flow is governed by the Stokes equation.
This is expected, because the Reynolds number $m\rho VH/\eta\approx 0.6$ 
for $\rho\approx 0.8/\sigma^3$, $V=0.005\sqrt{\epsilon/m}$,
$H\approx 300\sigma$, and $\eta\approx 2.0\sqrt{\epsilon m}/\sigma^2$.
Figure \ref{power-hyd} shows the slip profiles plotted on the log-log scale, 
which clearly exhibit the $1/x$ behavior extending from $|x|\approx 6l_s$ 
to $|x|\approx 270 l_s$. The inset to figure \ref{power-hyd} shows 
the scaled tangential velocity profiles at the solid surface, 
from which the existence of a universal slip profile is evident. 
Physically, when $H\gg l_s$, the Stokes flow is governed by 
only one velocity scale $V$ and one length scale $l_s$. Thus universality 
becomes evident from $v_x/V$ plotted as a function of $x/l_s$.  
A heuristic account of the universal slip profile is as follows.
Away from the MCL, the viscous shear stress is given by 
$-a\eta v_x(x)/|x|$, where $a$ is a constant $\sim 1$ and $v_x(x)$ is
the local tangential velocity. The NBC implies 
$v_x^{slip}(x)=-a l_s v_x(x)/|x|$. Combining this equation with 
$v_x^{slip}=v_x+V$ yields $v_x^{slip}(x)/V=1/(1+{|x|}/{al_s})$.
This relation, with $a\approx 2.14$ for best fit, agrees with 
the continuum slip profiles extremely well, as seen in the inset to 
figure \ref{power-hyd}.

\section{Scaling analysis: sharp or diffusive interface limit}\label{scaling}

In this section we look at the MCL problem from two different perspectives. 
These naturally arise from the physical reality that there are two distinct 
regions in the MCL problem: the interfacial region and the rest. 
In the perspective of the sharp interface limit, the problem is 
looked at from outside the interfacial region. The effect of 
the interfacial region is taken into account only in an integrated sense. 
In the diffuse interface limit, on the other hand, we examine the problem 
when the interfacial region is physically large. It turns out that 
the two limits can be mathematically analyzed by varying the magnitude 
of $M$ and $\Gamma$. In addition, we also give an abbreviated account
of the recent work by \cite{sharp-RE} on the sharp interface limit, 
which not only simplifies the mathematics, but 
also expresses the GNBC without the $\phi$ variable, and hence 
physically more transparent.

In appendix \ref{appendix-b} we have derived the total rate of energy 
dissipation in steady state (see equations (\ref{RDv}), (\ref{RRv}), 
and (\ref{dissipation-function-v})),
\begin{equation}\label{dissipation-function-vv}
\begin{array}{lll}
R_2\left[{\bf v}\right]
&=&R_v\left[{\bf v}\right]+R_s\left[{\bf v}\right]
+R_d\left[{\bf v}\right]+R_r\left[{\bf v}\right]\\
&=&\int d{\bf r}\left[\displaystyle\frac{\eta}{2}
(\partial_i v_j+\partial_j v_i)^2\right]+
\int dS\left[\beta\left(v_\tau^{slip}\right)^2\right]+\\
&&\int d{\bf r}_1d{\bf r}_2\left[-\displaystyle\frac{1}{M}
G({\bf r}_1,{\bf r}_2)\left({\bf v}\cdot\nabla\phi\right)_{{\bf r}_1}
\left({\bf v}\cdot\nabla\phi\right)_{{\bf r}_2}\right]+
\int dS\left[\displaystyle\frac{1}{\Gamma}
\left(v_\tau\partial_\tau\phi\right)^2\right].
\end{array}
\end{equation}
This function(al) is useful to the variational analysis of 
the sharp/diffuse interface limits. It is obtained by eliminating
the two rates $({\bf J},\dot{\phi})$ by expressing them in $\bf v$,
for the steady state. The Stokes equation (\ref{stokes}) and 
the GNBC (\ref{gnbc}) can be derived by 
minimizing $R_2[{\bf v}]$ with respect to $\bf v$,
supplemented with equations (\ref{advection-diffusion}) and
(\ref{relaxation}) for the linear dissipative dynamics of $\phi$
(see equations (\ref{ELeqB-app}) and (\ref{ELeqS-app})).

The CH diffuse-interface modeling allows diffusive transport through
the fluid-fluid interface (Seppecher 1996; Jacqmin 2000;
Chen, Jasnow \& Vinals 2000; Pismen \& Pomeau 2000;
Briant \& Yeomans 2004).
By taking the limit of $M\to 0$ and $\Gamma\to 0$, we obtain 
the sharp interface limit in which advection dominates such that 
$\dot{\phi}\to 0$ and 
${\partial\phi}/{\partial t}\to -{\bf v}\cdot\nabla\phi$.
According to equation (\ref{dissipation-function-vv}), 
if $1/M$ and $1/\Gamma$ both approach positive infinity, 
minimizing $R_2$ with respect to $\bf v$ then 
requires ${\bf v}\cdot\nabla\phi\to 0$ in the interfacial region
(see the last two terms on the right-hand side of 
equation (\ref{dissipation-function-vv})).
That means in steady state the flow is parallel to the interface. 

The limiting magnitude of ${\bf v}\cdot\nabla\phi$ can be obtained through 
a scaling analysis. We assume that the interfacial thickness $\xi$ is 
much smaller than the smallest interfacial curvature radius, a limit
realized for moderate shear rates. Integrating the Stokes equation 
(\ref{stokes}) across the fluid-fluid interface yields 
$|\mu|\phi_+\sim \eta||\nabla {\bf v}||$, from which we have 
$|\mu|\sim \eta V/l\phi_+$, where $V$ and $l$ are the characteristic 
velocity and length scales of the flow. Integrating 
equation (\ref{advection-diffusion}) (with ${\partial\phi}/{\partial t}=0$ 
in steady state) across the fluid-fluid interface yields
$|v_m|\phi_+\sim M||\nabla\mu||$, from which we have 
$|v_m|\sim M|\mu|/l\phi_+$, where $v_m$
is the velocity component along the interfacial normal $m$.
Together, $|\mu|\sim \eta V/l\phi_+$ and $|v_m|\sim M|\mu|/l\phi_+$ lead to
$|v_m|/V\sim M\eta/l^2\phi_+^2$ for the dimensionless interfacial 
normal velocity. This relation indicates that in order to realize
the sharp interface limit, $\sqrt{M\eta}/\phi_+$ is the length scale
which must be made small enough compared to $l$. 
(This length scale arises from the coupling of equations (\ref{stokes})
and (\ref{advection-diffusion}) (Bray 1994; Jacqmin 2000).)
As for the characteristic length $l$ of the flow field, 
it is given by the slip length $l_s$ in the inner region 
close to the contact line, where $l_s$ is actually 
the only length scale governing the Stokes flow. 
Far away from the contact line, $l$ becomes the dimension of 
the confined system, e.g., $l\sim H$. Typically $l_s$ is much smaller than
$H$, and consequently $|v_m|$ reaches the maximum in the immediate vicinity 
of the MCL. This leads to the first conclusion that 
the sharp interface limit is realized in the bulk when
\begin{equation}\label{normal-vel-B}
\displaystyle\frac{|v_m|}{V}\sim 
\displaystyle\frac{M\eta}{l_s^2\phi_+^2} \ll 1.
\end{equation} 
This relation ensures the flow to be parallel to the interface in 
the inner region of fast velocity variation and large interfacial curvature.

In the sharp interface limit, the tangential viscous stress is vanishingly 
small at the MCL because of the vanishing interfacial normal velocity there.
Here we consider the case of $\theta_s=90^\circ$ for simplicity and 
generalization to $\theta_s\ne 90^\circ$ is straightforward.
It follows that in steady state the near-complete slip at the MCL is mainly 
sustained by the uncompensated Young stress:
$\beta V\sim \gamma\cos\theta_d/\xi$, where $V$ is for the maximum slip 
velocity (the wall speed relative to the stationary fluid-fluid interface) 
and $\gamma\cos\theta_d/\xi$ is obtained by considering the integrated 
uncompensated Young stress in equation (\ref{unYS}) as distributed within 
$\xi$. The difference between the dynamic and static contact angles, 
$\Delta\theta=\theta_s-\theta_d$, is then obtained as
$\Delta\theta\sim \xi\beta V/\gamma=\xi Ca/l_s$,
where $l_s=\eta/\beta$ is the slip length and $Ca=\eta V/\gamma$ is 
the capillary number. 
Integrating equation (\ref{relaxation}) 
(with ${\partial\phi}/{\partial t}=0$)  
across the fluid-fluid interface along the solid surface yields 
the tangential velocity across the interface
$|v_\tau|\sim \Gamma K\cos\theta_d\approx \Gamma K\Delta\theta$.
(For $\theta_s=90^\circ$, the tangential direction $\tau$ is the same
as the interfacial normal $m$, and hence $v_\tau$ becomes $v_m$.)
Substituting $\Delta\theta\sim\xi Ca/l_s$ into
$|v_m|\sim\Gamma K\Delta\theta$, we obtain 
$|v_m|\sim\Gamma K\xi Ca/l_s$, and hence 
$|v_m|/V\sim \Gamma K\xi \eta/\gamma l_s$. 
Using $\gamma=2\sqrt{2}K\phi_+^2/3\xi$, we obtain
${|v_m|}/{V}\sim {\Gamma\xi^2\eta}/{\phi_+^2l_s}$. 
This leads to the second conclusion that the sharp interface limit 
is realized at the contact line when 
\begin{equation}\label{normal-vel-MCL}
\displaystyle\frac{|v_m|}{V}\sim 
\displaystyle\frac{\Gamma\xi^2\eta}{\phi_+^2l_s} \ll 1,
\end{equation}
which ensures the interface not to be penetrated by the flow. Note that 
the two conditions ({\ref{normal-vel-B}) and ({\ref{normal-vel-MCL})
are independent of the magnitude of $V$. This is consistent with 
the linearity of the hydrodynamic model (for sufficiently low flow rates) 
and has been well verified numerically.

In Sec. \ref{comparison}, it has been mentioned that $M$ and $\Gamma$ are 
treated as fitting parameters to optimize the MD-continuum comparison.
Since the fluid-fluid interface in our MD simulations is impenetrable,
the hydrodynamic model has to be within the sharp interface limit in order
to reproduce the MD results. It follows that the values of 
$M$ and $\Gamma$ must satisfy the conditions 
({\ref{normal-vel-B}) and ({\ref{normal-vel-MCL}). 
Moreover, as long as the sharp interface limit is reached, the continuum
predictions are not sensitive to the values of $M$ and $\Gamma$. 
Such has indeed been our experience. That means $M$ and $\Gamma$ should
not be regarded as fitting parameters in our MD-continuum comparison 
because they are simply used to realize the sharp interface limit,
in accordance with the interface impenetrability conditions
(viewed outside the interfacial region).

We have shown that in order to sustain the near-complete slip at the MCL,
the dynamic contact angle $\theta_d$ has to deviate from the static
angle $\theta_s$ by $\Delta\theta\sim \xi Ca/l_s$. 
A similar scaling relation has recently been obtained by \cite{sharp-RE}
in a general discussion on the sharp interface limit of the GNBC. 
Here we derive a complete expression for $\Delta\theta$ within 
the CH phase-field formulation.
Consider an interface deformed by the shearing movement of confining walls
(see figure \ref{md-geometry}b). In steady state the tangential viscous stress 
is negligibly small at the MCL where the slip is near-complete 
($v_x^{slip}\approx V$ at the lower fluid-solid interface). Therefore
the GNBC may be expressed as $\beta V=-K\partial_z\phi\partial_x\phi$.
Suppose the deviation of $\theta_d$ from $\theta_s=90^\circ$ is very small 
(this is indeed the case, as we will show later). 
Then $\partial_z\phi\approx -\partial_x\phi\cos\theta_d$ and
$\beta V\approx K\left(\partial_x\phi\right)^2\cos\theta_d$. 
Using the interfacial profile
$\phi(x)=\phi_{+}\tanh\left(x/{\sqrt{2}\xi}\right)$ at $z=0$
(for a gently deformed interface), we obtain
$\beta V\approx {K\phi_+^2}\cos\theta_d/{2\xi^2}$. 
It follows that
\begin{equation}\label{dyn-angle}
\Delta\theta=\displaystyle\frac{4\sqrt{2}}{3}
\displaystyle\frac{\xi}{l_s}\displaystyle\frac{\eta V}{\gamma}.
\end{equation}
This expression has been quantitatively verified.
Substituting the MD values $l_s=3.8\xi$, $\gamma=5.5\epsilon/\sigma^2$, 
$\eta=1.95\sqrt{\epsilon m}/\sigma^2$, and $V=0.25\sqrt{\epsilon/m}$ into 
equation (\ref{dyn-angle}) yields $\Delta\theta=0.044$ (or $2.5^\circ$), 
in excellent agreement with $\cos\theta_d=0.0437$ obtained from 
the $\phi=0$ locus in the continuum solution. 
Actually there is another measure of $\cos\theta_d$ using the integrated 
Young stress $\int_{int}dx K\partial_n\phi\partial_x\phi=
\gamma\cos\theta_d$, which produces a value only slightly different from that 
determined by the $\phi=0$ locus. We note that in our nanoscale MD simulations,
$l_s$ is typically larger than $\xi$ and $\gamma> 10\eta V$, and hence
$\Delta\theta <0.1$. Yet this small angle of deviation is necessary
for the uncompensated Young stress to sustain the near-complete slip 
of the MCL (Qian, Wang \& Sheng 2003).

In a recent study of the sharp interface limit of the GNBC 
by \cite{sharp-RE},
it has been shown that the deviation of the dynamic contact angle $\theta_d$
from the static contact angle $\theta_s$ is proportional to 
the dimensionless parameter $\beta^* V\delta/\gamma$, which measures
the relative strength of the frictional force between the fluid and the solid
and the interfacial force between the fluids. Here $\beta^*$ is 
the (average) slip coefficient in the contact line region, depending on
$\beta_1$ and $\beta_2$ in the two single-phase flow regions 
as well as the fluid-fluid interfacial structure, and $\delta$ is 
the fluid-fluid interfacial thickness. Numerical results have been obtained
for the relation between the microscopic contact angle $\theta_d$ 
and the apparent contact angle, demonstrating good agreement with 
the analytical results based on matched asymptotic expansions 
(Cox 1986).

From the above, it follows that in the sharp interface limit
the contact angle can be set at the value of the static contact angle
(for $\xi\to 0$ and/or $Ca\to 0$), 
and since outside the fluid-fluid interfacial region the NBC is valid, 
numerically the flow field can be calculated separately on the two sides of 
the interface, linked together via the interface transition relations 
(zero normal component of velocity, continuity of tangential
component of velocity, continuity of tangential stress, normal stress
difference across the interface being balanced by the tensile force
proportional to the interface curvature) 
(Zhou \& Sheng 1990; Ren \& E 2005b).  
However, in such numerical solutions the tangential (viscous) stress 
is necessarily discontinuous at the contact line. This can be clearly 
seen by considering the tangential viscous stresses approached along 
the three interfaces (two fluid-solid and one fluid-fluid) terminating
at the contact line. 
For simplicity, let us consider a fluid-fluid interface vertically 
intersecting the solid surface. Velocity continuity dictates that 
the tangential viscous stress at the solid surface approaches 
$\beta_1 V$ (or $\beta_2 V$) at the contact line in the single-phase 
flow region left (or right) to the fluid-fluid interface, 
with $\beta_1$ and $\beta_2$ being the slip coefficients
in the left and right regions, respectively. However, the impenetrability 
condition at the fluid-fluid interface dictates that along this interface, 
the velocity component parallel to the solid surface vanishes,
leading to a vanishing tangential viscous stress at the solid surface
when approached along the fluid-fluid interface down to the contact line. 
Hence there exist three distinct values for the tangential viscous stress.
The uncompensated Young stress thus enters as the required subsidiary
condition to complete the picture and make the solutions physically
meaningful.  
In particular, the uncompensated Young stress interpolates
between the two values of $\beta_1 V$ and $\beta_2 V$, with a mean value
given by $(\beta_1+\beta_2)V/2$, if the two fluids are assumed to interact
with the solid independently (Qian, Wang \& Sheng 2003).

Opposite to the sharp interface limit is the limit of $M\to\infty$ and 
$\Gamma\to\infty$. Obviously, in this limit minimizing $R_2$ in equation 
(\ref{dissipation-function-vv}) is equivalent to minimizing $R_1$ in 
equation (\ref{single-functional}) because $R_d$ and $R_r$ 
both vanish regardless of the velocity distribution 
(see equations (\ref{RDv}) and (\ref{RRv})). As a consequence, 
the flow field and the fluid-fluid interface are decoupled: 
the velocity is distributed as if there is only one single phase 
while the interfacial profile approaches the equilibrium one.
As the fluid-fluid interface becomes very transparent (through 
diffusive transport), the contact line loses its usual implications. 
Indeed, it has been shown that as this limit is approached, 
the stress singularity can be lifted even if the no-slip boundary 
condition is applied (Seppecher 1996; Jacqmin 2000;
Chen, Jasnow \& Vinals 2000; Briant \& Yeomans 2004). 
This result can be made intuitively plausible from our variational 
formulation as follows.

According to equation (\ref{dissipation-function-vv}), if the functional 
$R_2[{\bf v}]$ is to be minimized subject to $\beta$, $1/M$, and $1/\Gamma$ 
all approaching positive infinity, then the problem is reduced to 
solving the Stokes equation
subject to the no-slip boundary condition $v_\tau^{slip}=0$
and the interface impenetrability condition $v_m=0$. This leads to 
the well-known non-integrable singularity in viscous dissipation. 
Therefore, mathematically the contact-line singularity may be viewed as 
resulting from minimizing $R_2[{\bf v}]$ with $\beta$, $1/M$, and $1/\Gamma
\to\infty$. By removing either the $\beta\to\infty$ constraint (i.e.,
allowing slipping), or the $1/M$, $1/\Gamma\to\infty$ constraint
(i.e., allowing diffusive relaxation), the total dissipation can only 
decrease from infinity, thus regularizing the solution. 
This is especially the case since the divergence is logarithmic in
nature, i.e., the divergence is only marginal. 
We should note, however, that physically realistic cases correspond to
$\beta$, $1/M$, and $1/\Gamma$ all remain finite, as evidenced by MD results.
In fact, application of the above considerations to the problem of 
corner-flow singularity (involving a flow in a corner with one rigid 
plane sliding over another) (Batchelor 1991; Moffatt 1964;
Koplik \& Banavar 1995) would be equally valid (Qian \& Wang 2005).

It is interesting to note that in either of the two limits discussed above,
the rate of interfacial dissipation, $R_d+R_r$, tends to vanish.
In the sharp interface limit of $M\to 0$ and $\Gamma\to 0$, 
the limiting behaviors of the interfacial normal velocity expressed in
equations (\ref{normal-vel-B}) and (\ref{normal-vel-MCL}) make 
$R_d\sim M$ and $R_r\sim \Gamma$.
(Equation (\ref{RDv}) indicates $R_d\propto |v_m|^2/M$ while 
equation (\ref{normal-vel-B}) indicates $|v_m|\propto M$, and hence
$R_d\propto M\to 0$. Similarly, $R_r\propto \Gamma\to 0$.)
In the opposite limit of $M\to\infty$ and $\Gamma\to\infty$,
as the interface is penetrated by the flow, 
$R_d$ and $R_r$ simply vanish as $1/M$ and $1/\Gamma$, respectively.
That the positive definite rate of interfacial dissipation, 
$R_d+R_r$, approaches zero in the two opposite limits implies that
a maximum should be reached somewhere in between. 
However, the total rate of dissipation $R_2$ should increase monotonically
from the limit of $M\to\infty$ and $\Gamma\to\infty$ to that of
$M\to 0$ and $\Gamma\to 0$. Although there is only $R_v+R_s$ left for
$R_2$ in either of the two limits, the latter (sharp interface) limit 
imposes vanishing interfacial normal velocity as the additional condition.
This would certainly lead to a flow whose total rate of dissipation is 
larger than that obtained from minimizing the same functional without 
the additional constraint.

We want to point out that while mathematically the sharp interface limit
may be simply obtained by excluding $R_d$ and $R_r$ from $R_2$ and
applying $v_m=0$ at the interface instead, physically the two-phase 
interfacial dissipation may not always be negligible, especially when 
the interfacial region of partial miscibility has finite and non-negligible
width such that structural relaxation may occur.

It should be noted that dimensional analysis indicates that 
$\left[M\right]=\left[\Gamma\right]\left[{\rm Length}\right]^3$, i.e., 
there is a length scale $l_0$ which links these two parameters 
through the relation $M=\Gamma l_0^3$. Physically it is plausible to assume 
that $l_0$ is determined by a combination of microscopic factors, such as 
fluid-fluid interaction, fluid-solid interaction, molecular organization 
of the fluids, and molecular structure of the wall.
Hence $M$ and $\Gamma$ may be physically related in any given system.

\section{Concluding remarks}\label{remarks}

We should point out the inadequacies in our present formulation. First,
the free energy used to delineate the two fluids is a minimal model.
It neglects, for example, the density variation that can be fairly 
significant in the interfacial region. Second, our model in its present
form is only applicable to simple liquids. Complex fluids would require 
nontrivial extensions. Third, we have neglected the van der Waals interaction
which is very important in understanding precursor films. These and other
inadequacies represent tasks still to be pursued. 
The main purpose of this paper is to outline the framework of 
a general theory which can resolve the MCL problem in its simplest form.

It is important to emphasize that while the partial slip in single-phase 
flows is generally small and quantitatively indistinguishable from no-slip, 
yet its significance is qualitatively much greater.  First, it is clear 
from the above that the GNBC goes hand-in-hand with the NBC in single-phase 
flows, and that it is incompatible with the no-slip boundary condition even 
in single-phase flows. The latter is clear from the power-law partial slip 
which extends mesoscopic distances into single-phase flow regimes.  
Second, even if the slip is small, the fact that slip exists means that 
its magnitude may be manipulated, i.e., it can be made larger or smaller.  
In particular, since the slip coefficient is a thermodynamic quantity, 
just as the viscosity, its magnitude should depend on molecular 
interactions and interface geometries (as well as the state variables 
such as temperature) (Barrat \& Bocquet 1999b; Leger 2003;
Granick, Zhu \& Lee 2003; Zhu \& Granick 2004; Neto \etal 2005), 
a fact which can be used to advantage 
experimentally through nanoscale manipulations and environmental controls.
For example, effective slip at nano-patterned surfaces has already been 
studied (Philip 1972; Lauga \& Stone 2003;
Cottin-Bizonne \etal 2003; Cottin-Bizonne \etal 2004;
Priezjev, Darhuber \& Troian 2005; Qian, Wang \& Sheng 2005). 
In contrast, no-slip boundary condition is 
a clean-cut statement, with no room for adjustment or 
for physics considerations, only carries with it the burden of proof.  
Here the broad applicability of the no-slip boundary condition can not be 
considered as proof against slipping, as a very small amount of partial slip
would clearly lead to similar results. 
It is therefore rather obvious that whereas slip/partial-slip can be 
derived from general principles and demonstrated through MD simulations, 
no-slip has yet to be proved with similar generalities.     
In fact, at present slip is already a subject with some fairly extensive 
literature. We refer to the reviews by \cite{granick1} and 
by \cite{exp-review} for a more complete list of references.

 In closing, we note that in the case of the moving contact line, 
complete slip occurs in the linear regime (i.e., $\beta$ is a constant).  
This is in contrast to the view that complete slip can only occur 
when ``interface fracturing'' occurs, i.e., in the fully nonlinear regime 
(sometimes also denoted as ``super slipping'' or ``threshold slipping''), 
corresponding to a stress-dependent $\beta$ 
(see e.g. Thompson \& Troian 1997).  
It is rather likely that a statistical mechanical study of 
the slip coefficient can indeed produce a threshold behavior, e.g., 
$\beta$ approaches zero as the shear stress exceeds 
a certain threshold.  However, in the linear regime this is not the case.  
The difference between the MCL complete slip and the interface fracturing
is that in the case of MCL, the (complete) slip velocity can be 
very small and the relevant shear stress can be low.  The fact that 
complete slip can occur in this case is due to the localized nature of 
the uncompensated Young stress, in addition to the tangential viscous stress.  
Thus we can have complete slip in both the linear and nonlinear regimes, 
with different underlying physics.

The authors are grateful to Weinan E, Chun Liu and Weiqing Ren for 
helpful discussions.
This work was partially supported by the Hong Kong RGC CERG 
No. 604803, the RGC central allocation grant CA05/06.SC01, and
the Croucher Foundation Grant Z0138.

\appendix
\section{The principle of minimum energy dissipation}\label{appendix-a}

To outline the principle of minimum energy dissipation 
(Onsager 1931a and 1931b), 
consider a system described by one single variable $\alpha$,
governed by the overdamped Langevin equation
\begin{equation}\label{langevin}
\gamma\dot{\alpha}=-\displaystyle\frac{\partial F(\alpha)}{\partial \alpha}
+\zeta(t),
\end{equation}
where $\gamma$ is the frictional coefficient, $\dot{\alpha}$ is the rate 
of change of $\alpha$, $F(\alpha)$ is the free energy,
and $\zeta(t)$ is a white noise satisfying
$\langle\zeta(t)\zeta(t')\rangle=2\gamma k_BT\delta(t-t')$, 
with $k_B$ denoting the Boltzmann constant and $T$ the temperature.
The probability of finding the system in the state described by $\alpha$ is
a function of time, denoted by $P(\alpha,t)$ and governed by 
the Fokker-Planck equation 
\begin{equation}\label{fokker-Planck}
\displaystyle\frac{\partial P}{\partial t}=D\left[\displaystyle\frac
{\partial^2 P}{\partial \alpha^2}+\displaystyle\frac{1}{k_BT}
\displaystyle\frac{\partial}{\partial \alpha}\left(\displaystyle\frac
{\partial F}{\partial \alpha}P\right)\right],
\end{equation}
where $D$ is the diffusion coefficient satisfying the Einstein relation
$\gamma D=k_BT$. It is clear that the Boltzmann distribution 
$P_{eq}(\alpha)\propto\exp\left[-F(\alpha)/k_BT\right]$ is a stationary 
solution of the Fokker-Planck equation (\ref{fokker-Planck}).
It can be shown (Langer 1968) that the transition probability from
$\alpha$ at $t$ to $\alpha'$ at $t+\Delta t$, i.e., 
$P_2(\alpha',t+\Delta t;\alpha,t)$, is given by
\begin{equation}\label{transition-probability}
P_2(\alpha',t+\Delta t;\alpha,t)=\displaystyle\frac{1}
{\sqrt{4\pi D\Delta t}}
\exp\left[-\displaystyle\frac{(\alpha'-\alpha)^2}{4D\Delta t}\right]
\exp\left[-\displaystyle\frac{F(\alpha')-F(\alpha)}{2k_BT}\right],
\end{equation}
for $\alpha'$ in the vicinity of $\alpha$ and short time interval 
$\Delta t$. 
The most probable transition occurs between $\alpha$ and $\alpha'$ is 
the one which minimizes
\begin{equation}\label{transition-condition}
A=\displaystyle\frac{\gamma(\alpha'-\alpha)^2}{2\Delta t}
+\left[F(\alpha')-F(\alpha)\right]\approx\left[\displaystyle\frac
{\gamma}{2}\dot{\alpha}^2+\displaystyle\frac
{\partial F(\alpha)}{\partial \alpha}\dot{\alpha}\right]\Delta t,
\end{equation}
Here $\dot{\alpha}=(\alpha'-\alpha)/\Delta t$ and the minimum of $A$ 
is taken with respect to $\alpha'$, or equivalently $\dot{\alpha}$, 
for prescribed $\alpha$. 
The Euler-Lagrange equation for minimizing $A$ is thus
\begin{equation}\label{transition-ELeq}
\gamma\dot{\alpha}=\displaystyle\frac{\gamma(\alpha'-\alpha)}{\Delta t}
=-\displaystyle\frac{\partial F(\alpha)}{\partial \alpha},
\end{equation}
as expected from the Langevin equation (\ref{langevin}). 
Equation (\ref{transition-ELeq}) is actually the simplest, 
one-variable version of the linear relation 
in equation (\ref{onsager-1}) for rates and forces, and the function 
$\gamma\dot{\alpha}^2/2+(\partial F/\partial \alpha)\dot{\alpha}$
in $A$ is the corresponding one-variable version of the function
$\Phi(\dot{\alpha},\dot{\alpha})+\dot{F}(\alpha,\dot{\alpha})$
in Onsager's variational principle, stated by
equations (\ref{onsager-2}), (\ref{onsager-3}), and (\ref{onsager-4}).
From the above discussion, it is clear that (1) the variation of
$A$ should be taken with respect to the rate $\dot{\alpha}$, for prescribed
state variable $\alpha$, (2) the minimum dissipation principle implies
the balance of dissipative force and the force derived from free energy
(see equation (\ref{transition-ELeq})), and (3) the minimum dissipation
principle yields the most probable course of a dissipative process,
provided the displacement from the equilibrium is small (Onsager 1931b,
Onsager \& Machlup 1953).

\section{The dissipation function in steady state}\label{appendix-b}

In a steady state with ${\partial\phi}/{\partial t}=0$, 
the total rate of energy dissipation $R_2=2\Phi$ in equations 
(\ref{dissipation-function-1}) and (\ref{dissipation-function-2})
can be expressed as a functional of ${\bf v}({\bf r})$ only. 
This is because for prescribed phase field $\phi$, 
the rate $\bf J$ in the bulk can be determined from
$\dot{\phi}={\bf v}\cdot\nabla\phi$ through equations 
(\ref{diffusive-current}) and (\ref{advection-diffusion}) 
while the rate $\dot{\phi}=v_\tau\partial_\tau\phi$ 
at the solid surface is already given in terms of ${\bf v}$. 
By expressing $R_2$ as a functional of ${\bf v}({\bf r})$, we can obtain 
a form of the dissipation function that is useful to the variational 
analysis of the sharp interface limit and the diffuse interface limit.

From equation (\ref{advection-diffusion}) 
with ${\partial\phi}/{\partial t}=0$, 
we can formally express $\mu({\bf r})$ by
\begin{equation}\label{mu-Green}
\mu({\bf r})=\displaystyle\frac{1}{M}\int d{\bf r}'G({\bf r},{\bf r}')
\left({\bf v}\cdot\nabla\phi\right)_{{\bf r}'},
\end{equation}
where $G({\bf r},{\bf r}')$ is the Green function 
for the Laplacian operator satisfying the boundary condition 
$\partial_n\mu=0$ at the solid surface.
From $R_d$ in equation (\ref{RD0}) and $J=-M\nabla\mu$ 
(equation (\ref{diffusive-current})), we have
\begin{equation}\label{RDv0}
R_d=\int d{\bf r}\left[M(\nabla\mu)^2\right]=
\int d{\bf r}\left[-M\mu\nabla^2\mu\right],
\end{equation}
where the integration by parts has been used with $\partial_n\mu=0$ 
at the solid surface. 
Substituting equation (\ref{mu-Green}) into (\ref{RDv0}) yields
\begin{equation}\label{RDv}
R_d\left[{\bf v}\right]
= \int d{\bf r}_1d{\bf r}_2\left[-\displaystyle\frac{1}{M}
G({\bf r}_1,{\bf r}_2)\left({\bf v}\cdot\nabla\phi\right)_{{\bf r}_1}
\left({\bf v}\cdot\nabla\phi\right)_{{\bf r}_2}\right].
\end{equation}
Meanwhile, substituting $\dot{\phi}=v_\tau\partial_\tau\phi$
into $R_r$ in equation (\ref{RR0}), we obtain
\begin{equation}\label{RRv}
R_r\left[{\bf v}\right]= \int dS\left[\displaystyle\frac{1}{\Gamma}
\left(v_\tau\partial_\tau\phi\right)^2\right].
\end{equation}
Combining equations (\ref{RV}), (\ref{RS}), (\ref{RDv}), and (\ref{RRv}) 
with $R_2$ defined in equations (\ref{dissipation-function-1}) 
and (\ref{dissipation-function-2}), we obtain 
\begin{equation}\label{dissipation-function-v}
\begin{array}{lll}
R_2\left[{\bf v}\right]
&=&R_v\left[{\bf v}\right]+R_s\left[{\bf v}\right]
+R_d\left[{\bf v}\right]+R_r\left[{\bf v}\right]\\
&=&\int d{\bf r}\left[\displaystyle\frac{\eta}{2}
(\partial_i v_j+\partial_j v_i)^2\right]+
\int dS\left[\beta\left(v_\tau^{slip}\right)^2\right]+\\
&&\int d{\bf r}_1d{\bf r}_2\left[-\displaystyle\frac{1}{M}
G({\bf r}_1,{\bf r}_2)\left({\bf v}\cdot\nabla\phi\right)_{{\bf r}_1}
\left({\bf v}\cdot\nabla\phi\right)_{{\bf r}_2}\right]+
\int dS\left[\displaystyle\frac{1}{\Gamma}
\left(v_\tau\partial_\tau\phi\right)^2\right],
\end{array}
\end{equation}
for the total rate of dissipation in two-phase flows,
which is the sum of $R_v$ due to viscosity, 
$R_s$ due to slipping, $R_d$ due to diffusion in the bulk, 
and $R_r$ due to relaxation at the surface.

In accordance with the principle of minimum energy dissipation
(Onsager 1931a and 1931b), for steady state the equation(s) of motion 
can be derived from minimizing the dissipation-function 
$\Phi$ ($=\frac{1}{2}R_2$ here) with respect to the rates. 
Here we note that in steady state, the rates $\bf J$ in the bulk and
$\dot{\phi}$ at the solid surface are already determined by $\bf v$
for prescribed $\phi$, and $R_2[{\bf v}]$ in equation 
(\ref{dissipation-function-v}) displays a symmetric, quadratic form 
as a function(al) of $\bf v$, in accordance with the reciprocal relations,
i.e., $\rho_{ij}=\rho_{ji}$ for the coefficients $\rho_{ij}$
in $\Phi(\dot{\alpha},\dot{\alpha})=\frac{1}{2}\sum_{i,j}\rho_{ij}
\dot{\alpha}_i\dot{\alpha}_j$ (equation (\ref{onsager-3})).
The variation of the total dissipation $R_2[{\bf v}]$ should be taken 
with respect to $\bf v$ only, as it is the only rate.

Based on equation (\ref{dissipation-function-v}), 
minimizing $R_2$ with respect to ${\bf v}$ 
in the bulk yields the Stokes equation (\ref{stokes}), while
minimizing $R_2$ with respect to tangential fluid velocity $v_\tau$ 
at the solid surface yields the GNBC (\ref{gnbc}). That is,
consider a variation of the velocity field ${\bf v}({\bf r})\to 
{\bf v}({\bf r})+\delta{\bf v}({\bf r})$. The associated changes
in $R_v$ and $R_s$ are already given by equations (\ref{var-Rv}) and 
(\ref{var-Rs}), and those in $R_d[{\bf v}]$ and $R_r[{\bf v}]$ are given by
\begin{equation}\label{var-Rd}
\delta R_d=\int d{\bf r}d{\bf r}'
\left[-\displaystyle\frac{2G({\bf r},{\bf r}')
\left({\bf v}\cdot\nabla\phi\right)_{{\bf r}'}
\left(\delta{\bf v}\cdot\nabla\phi\right)_{{\bf r}}}{M}\right]
=-2\int d{\bf r}\left[\mu\partial_i\phi\delta v_i\right],
\end{equation}
and
\begin{equation}\label{var-Rr}
\delta R_r=\int dS\left[\displaystyle\frac{2}{\Gamma}
\left(v_\tau\partial_\tau\phi\right)\delta v_\tau\partial_\tau\phi\right]
=-2\int dS\left[L(\phi)\partial_\tau\phi\delta v_\tau\right],
\end{equation}
where equations (\ref{advection-diffusion}) and (\ref{relaxation}) 
have been used.
Combining equations (\ref{var-Rv}), (\ref{var-Rs}), (\ref{var-constraint}),
(\ref{var-Rd}), and (\ref{var-Rr}), we obtain the Euler-Lagrange equations
\begin{equation}\label{ELeqB-app}
-2\eta\partial_j\left(\partial_j v_i+\partial_i v_j\right)
-\partial_i\alpha-2\mu\partial_i\phi=0
\end{equation}
in the bulk and 
\begin{equation}\label{ELeqS-app}
2\eta\left(\partial_n v_\tau+\partial_\tau v_n\right)+2\beta v_\tau^{slip}
-2L(\phi)\partial_\tau\phi=0
\end{equation}
at the surface.
Note that equation (\ref{ELeqB-app}) is identical to the Stokes equation 
(\ref{stokes}) with $\partial_j v_j=0$ and $\alpha=-2p$, 
and equation (\ref{ELeqS-app}) reduces to the GNBC (\ref{gnbc}).

\end{document}